\documentclass[11pt]{article}

\usepackage[final]{acl}

\usepackage{times}
\usepackage{latexsym}

\usepackage[T1]{fontenc}

\usepackage[utf8]{inputenc}

\usepackage{microtype}

\usepackage{inconsolata}

\usepackage{graphicx}

\usepackage{listings,xcolor}
\lstdefinestyle{prompt}{
  basicstyle=\ttfamily\footnotesize, breaklines=true, frame=single,
  backgroundcolor=\color{gray!8}, columns=fullflexible, showstringspaces=false,
  xleftmargin=4pt, xrightmargin=4pt, aboveskip=4pt, belowskip=4pt}

\usepackage{url}
\usepackage{booktabs,multirow}
\usepackage{array}
\usepackage{longtable}
\usepackage{makecell} 

\title{Do LLMs Change Their Minds Like Humans? Diagnosing Human--LLM Divergence in Single-Turn Persuasion Judgments}

\author{
 \textbf{Lin Chen\textsuperscript{1,2}},
 \textbf{Yitong Chen}\textsuperscript{3},
 \textbf{Yong Li}\textsuperscript{$\dagger$3}
\\
 \textsuperscript{1}Network Science Institute, Northeastern University \\
 \textsuperscript{2}Department of Physics, Northeastern University \\
 \textsuperscript{3}Department of Electronic Engineering, BNRist, Tsinghua University \\
\small{$\dagger$ Corresponding author.}
 \\
 \small{
   \textbf{Correspondence:} \href{mailto:liyong07@tsinghua.edu.cn}{liyong07@tsinghua.edu.cn}
 }
}

\begin{document}
\maketitle

\begin{abstract}

Large language models (LLMs) are increasingly deployed as proxies for human participants in social simulations, yet whether they update their beliefs in response to persuasive arguments, as humans do, remains poorly understood. 
We conduct a systematic comparison using a naturally occurring online persuasion corpus in which original posters explicitly verify whether a reply changed their view.
Our results show that LLMs achieve only slight agreement with humans (Cohen's $\kappa$ ranging from 0.079 to 0.178). 
Content-level analyses show that humans and LLMs agree on the strongest persuasion cues but diverge on finer ones: humans are more swayed by novel content and assertive language, whereas LLMs favor topical similarity and surface-level formatting.
At the level of persuasion strategy, LLMs underweight emotional appeals and overweight credibility signals relative to humans, while the type of proposition under debate exerts no measurable effect on the degree of divergence. 
Furthermore, switching from first-person role-playing to third-person observation shifts all models toward greater resistance to persuasion, with the effect varying across persuasion strategies and textual features.
These findings highlight the risk of treating LLM judgments as faithful proxies for human belief updating and point to structural differences in how LLMs and humans process persuasive discourse.
Our code is available at \url{https://github.com/tsinghua-fib-lab/LLM-belief-update-cmv}.




\end{abstract}

\section{Introduction}

Large language models (LLMs) are increasingly used for simulating human interactions in various contexts~\cite{park2023generative,argyle2023out,gao2024large}, including online discourse~\cite{chuang2024simulating}, political elections~\cite{zhang2024electionsim}, and collective decision-making~\cite{jarrett2025language}.
A key cognitive process in these interactions is belief updating~\cite{anderson1981foundations,hogarth1992order}, through which individuals selectively revise their prior beliefs after encountering new evidence or arguments.
When exposed to the same arguments that humans find persuasive or unpersuasive, do LLMs revise or maintain their positions in a similar manner? 
If systematic divergence exists, applications that assume human-like reasoning risk producing distorted outcomes~\cite{chen2026ai,anthis2025llm}. 
Therefore, to achieve simulations with fidelity, it is critical to understand whether such divergence exists, and if so, what modulates it.

Prior work has debated whether LLMs can faithfully reproduce human opinion distributions, with evidence of demographic-level fidelity on survey tasks~\cite{argyle2023out} but also substantial misalignment for specific subgroups and topics~\cite{santurkar2023whose}. 
In argumentative settings specifically, studies have documented accuracy bias driving group  consensus~\cite{chuang2024simulating}, inherent social biases overriding assigned ideological perspectives in political debates~\cite{taubenfeld2024systematic}, and context accumulation shifting LLM judgments in predictable ways~\cite{geng2025accumulating}. 
Although these studies characterize how LLMs behave in argumentative contexts, they do not benchmark LLM responses against human counterparts for individual arguments, or examine what content-level properties of the argument, such as persuasion strategy, proposition type, or textual features, modulate LLM–human divergence. 
Moreover, recent work has shown that assigning actor versus observer roles to LLM agents induces systematic attribution biases~\cite{li2026taming}, but whether a perspective effect extends to belief updating remains untested.

Motivated by these research gaps, we ask:
\begin{itemize}
    \item \textbf{RQ1}: Do LLMs exhibit systematic bias in their belief update judgments relative to humans?
    \item \textbf{RQ2}: Do characteristics of the discussion content, including proposition type, persuasion strategy, and textual features, modulate the degree of this bias?
    \item \textbf{RQ3}: Does first-person role-playing introduce additional bias compared to third-person observation?
\end{itemize}

To answer these questions, we design a systematic comparison based on the \textit{ChangeMyView (CMV)} corpus~\cite{tan2016winning}, which provides naturally occurring, participant-verified labels of persuasion success. 
From \textit{CMV} discussion trees, we extract matched pairs of replies to the same original post, where one reply was acknowledged by the original poster as having changed their view and the other was not. 
We then present each reply independently to eight LLMs and ask them to judge whether the reply would change the view expressed in the original post. 

Our results reveal significant inconsistencies between human and LLM belief updates.
Across eight tested LLMs, belief update judgments under the first-person condition achieve only slight agreement with human labels, with models splitting into resistant, receptive, and neutral error profiles. 
Content-level analyses reveal that LLMs rely on a different set of textual cues than humans, favoring topical similarity over novelty and showing no sensitivity to structured formatting. 
At the level of persuasion strategy, LLMs underweight emotional appeals and overweight credibility signals relative to humans, while proposition type has no measurable effect on the degree of divergence.
Furthermore, switching from first-person role-playing to third-person observation shifts all models toward greater resistance to persuasion, with the effect varying across persuasion strategies and textual features.

Our contributions are summarized as follows: 

\begin{itemize}
    \item We present the first systematic comparison of LLM belief update judgments against human labels on real-world single-round persuasion data, and find that all models achieve only slight agreement with humans.
    \item We introduce proposition type and persuasion strategy as analytical dimensions for diagnosing LLM belief update behavior, and design automated annotation schemes. 
    \item We show that the human--LLM divergence is modulated by how the argument is written and what persuasion strategy it employs, but not by the type of claim being debated.
    \item We identify the role of perspective framing (first-person role-playing vs. third-person observation) in shaping LLM belief update behavior, showing that it systematically alters error composition, textual feature sensitivities, and responsiveness to different persuasion strategies.
\end{itemize}


\section{Experimental Design}

In this section, we present our experimental design around two components.
We first describe the dataset (Section~\ref{sec::dataset_construction}) and task formulation (Section~\ref{sec:task_formation}) used to elicit LLM belief update judgments, then introduce the three analytical dimensions along which we diagnose LLM–human divergence: textual features of the post and reply (Section~\ref{sec::extracting_textual_features}), proposition type of the original post (Section~\ref{sec::proposition_type_annotation}), and persuasion strategy of the challenger reply (Section~\ref{sec::persuasion_strategy_annotation}).

\subsection{Dataset Construction} \label{sec::dataset_construction}

We construct our dataset from the \textit{ChangeMyView (CMV)} corpus compiled by \citet{tan2016winning}.
CMV is an English-dominant Reddit community in which users publish posts expressing a personal view together with the reasoning behind it, and invite others to challenge that view. When a reply succeeds in shifting the original poster's (\textit{OP}'s) perspective, the \textit{OP} explicitly acknowledges this by replying with a delta ($\Delta$) character.  

From the CMV discussion trees, we first identify original posts that contain at least one root reply directly awarded a delta, excluding cases where the delta was awarded only after subsequent back-and-forth exchanges. 
This restriction ensures that each positive label reflects the persuasive effect of a single argument rather than the cumulative influence of an extended dialogue. 
This filtering yields 2,262 original posts. 
For each successful root reply, we then randomly pair it with one unsuccessful root reply from the same discussion tree, producing 3,016 matched pairs. 
Each pair shares the same original post but differs in persuasion outcome.



\subsection{Task Formation} \label{sec:task_formation}

We evaluate LLMs' belief update behavior through an independent judgment task: given an original post and a challenger reply, the model judges whether the reply would change the view expressed in the original post, outputting a binary decision (\texttt{delta\_awarded}: true or false). 

To test whether active stance maintenance introduces additional bias, we instantiate this task under two perspective conditions:
\begin{itemize}
    \item First-person: the model is instructed to role-play as the original poster and judge whether the reply has changed ``its own'' view (see Appendix~\ref{app::first_person_prompt} for detailed prompts).
    \item Third-person: the model acts as an external observer and predicts whether the reply would have changed the original poster's view (see Appendix~\ref{app::third_person_prompt} for detailed prompts).
\end{itemize}

To quantify the alignment between LLM judgments and human labels, we treat the human delta label as the ground truth, and report Cohen's $\kappa$, which corrects for chance agreement. 
We additionally report the True Positive Rate (TPR) and False Positive Rate (FPR) to reveal the directional structure of any bias: a high FPR indicates over-acceptance of ineffective arguments, while a low TPR indicates systematic rejection of effective ones.

\subsection{Extracting Textual Features} \label{sec::extracting_textual_features}

To examine whether surface-level textual properties of the original post and the reply predict persuasion outcomes differently for humans and LLMs, we extract nine features from each discussion pair.
On the original post side, to capture the volume of the initial argument, we measure post length (\textit{OP Length}). 
To gauge how personally invested the poster is in the claim~\cite{pennebaker2011secret}, we measure the frequency of first-person singular pronouns such as "I" and "my" (\textit{OP 1stPerson}). 
To reflect the rigidity of the stated position, we measure the frequency of definitive expressions such as "certainly," "nothing," and "anyone" (\textit{OP Definitive}). 
On the reply side, we measure reply length (\textit{Reply Length}) and the frequency of definitive expressions (\textit{Reply Definitive}) as counterparts to the corresponding OP features. 
To capture inclusive framing that appeals to shared identity, we measure the frequency of first-person plural pronouns such as "we" and "our" (\textit{Reply Inclusive}). 
To assess the presentational effort of the reply to reduce cognitive load~\cite{lorch1989text}, we record whether it contains formatting such as bullet points or numbered lists (\textit{Reply Formatting}) and whether it includes external links (\textit{Reply Link}). 
Finally, to capture the extent to which the reply introduces novel content beyond the scope of the original discussion~\cite{tan2016winning}, we compute the lexical dissimilarity between the two texts (\textit{Reply Dissimilarity}), measured by one minus the Jaccard coefficient over non-stopword vocabularies.

\subsection{Proposition Type Annotation} \label{sec::proposition_type_annotation}

To examine whether the degree of LLM–human divergence varies across discussion topics, we use an LLM-as-a-judge approach to classify the core claim of each original post into one of three proposition types following the classical taxonomy of proposition types in argumentation and debate studies~\cite{freeley2009argumentation,rybacki2008advocacy}:

\begin{itemize}
    \item Propositions of \textit{fact}, which assert empirical claims about the state of the world and whose validity could in principle be adjudicated through evidence, data, or logical demonstration; 
    \item Propositions of \textit{value}, which render evaluative judgments on the merit, morality, or importance of a subject, with the disagreement centering on normative criteria rather than empirical facts; 
    \item Propositions of \textit{policy}, which advocate for or against a specific course of action, law, or behavioral norm.
\end{itemize}

We adopt a single-label scheme in which each post is assigned to one and only one category that best characterizes the claim that the post is ultimately defending (see Appendix~\ref{app::proposition_type_annotation_prompt} for detailed prompts). 

\subsection{Persuasion Strategy Annotation} \label{sec::persuasion_strategy_annotation}

To examine whether different persuasion strategies employed in a reply influence the degree of LLM–human divergence, we also employ an LLM judge to annotate each challenger reply according to the three classical modes of persuasion~\cite{hidey2017analyzing,wachsmuth2017computational}:

\begin{itemize}
    \item \textit{Logos}, which appeals to reason through logical argument, factual evidence, statistics, or causal reasoning; 
    \item \textit{Pathos}, which appeals to the audience's emotions, empathy, or sense of identification; 
    \item \textit{Ethos}, which appeals to credibility established through personal experience, professional expertise, or reference to authoritative sources.
\end{itemize}

Different from topic classification, we adopt a multi-label binary scheme: each mode is independently judged as present or absent, since a single reply may combine multiple strategies (e.g., citing statistics while also sharing a personal story to evoke empathy) (see Appendix~\ref{app::strategy_annotation_prompt} for detailed prompts).

\subsection{Experimental Setup}

We test eight models spanning different model families and language ecosystems: \texttt{DeepSeek-V3}, \texttt{MiniMax-M2.5}, \texttt{GLM-4.7}, \texttt{GPT-4o-mini}, \texttt{GPT-5.5}, \texttt{Qwen2.5-32B-Instruct}, \texttt{Qwen2.5-72B-Instruct}, and \texttt{Gemini-2.5-Flash-Lite}.
For brevity, we refer to the last three in later sections as \texttt{Qwen-32B}, \texttt{Qwen-72B}, and \texttt{Gemini-2.5-Flash}, respectively.
All experiments are conducted with temperature = 0.1 (see Appendix~\ref{app::stability_across_runs} for stability analysis).

For both proposition type annotation and persuasion strategy annotation, we use  \texttt{GPT-5.1}, because it does not appear in our experimental model set, thereby avoiding circularity between the annotation process and the phenomena under investigation (see Appendix~\ref{app::annotation_validation} for validation against human annotations and another LLM judge).

\section{LLM-Human Divergence in Belief-Update Judgments (RQ1)} \label{sec::RQ1}

\begin{figure}[ht]
    \centering
    \includegraphics[width=\linewidth]{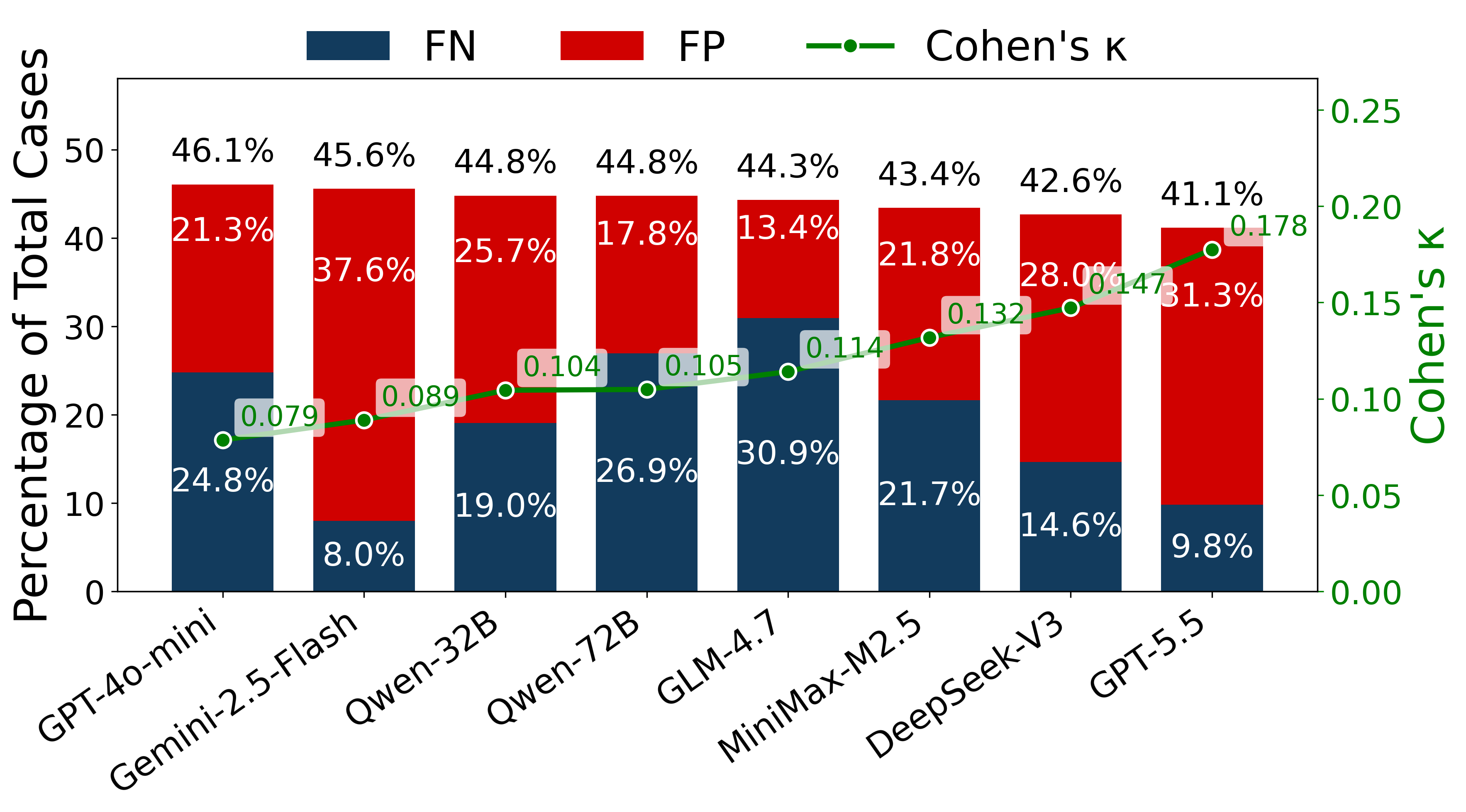}
    \caption{Discrepancy between LLMs and humans in belief update judgments. LLMs are sorted by their Cohen's $\kappa$ consistency with the human counterpart.}
    \label{fig:anomalies_and_kappa}
\end{figure}

To quantify LLM-human divergence in belief update judgments, we evaluate all eight models under the first-person condition described in Section~\ref{sec:task_formation} and compare their binary outputs against human labels.
As Figure~\ref{fig:anomalies_and_kappa} shows, Cohen's $\kappa$ between human and LLM judgments ranges from 0.079 (\texttt{GPT-4o-mini}) to 0.178 (\texttt{GPT-5.5}), all falling within the ``slight'' agreement category~\cite{landis1977measurement}.
In other words, even the best-performing model captures less than one fifth of the non-chance agreement structure, indicating that LLM belief update judgments consistently fail to align with human labels.

Across models, the total anomaly rates (FN + FP as a proportion of all judgments) are remarkably stable, spanning a narrow range from 41.1\% to 46.1\%.
However, the internal composition of these errors varies substantially across models and falls into three profiles. 
\texttt{GPT-4o-mini}, \texttt{Qwen-72B}, and \texttt{GLM-4.7} exhibit an \textit{resistant} profile (FN > FP), in which the model tends to maintain its assigned position and reject incoming arguments; \texttt{GLM-4.7} shows the most extreme split (FN = 30.9\%, FP = 13.4\%).
\texttt{Gemini-2.5-Flash}, \texttt{Qwen-32B}, \texttt{DeepSeek-V3}, and \texttt{GPT-5.5} exhibit a \textit{receptive} profile (FP > FN), in which the model tends to accept incoming arguments regardless of whether humans found them persuasive, with \texttt{Gemini-2.5-Flash} at the extreme (FN = 8.0\%, FP = 37.6\%).
Finally, \texttt{MiniMax-M2.5} exhibits a \textit{neutral} profile with nearly symmetric errors (FN = 21.7\%, FP = 21.8\%). 
Nevertheless, different LLMs agree with one another considerably more than any of them agrees with humans (Appendix~\ref{app::consistency_across_models}), suggesting that a shared LLM-specific judgment structure underlies the surface-level variation across models.

In short, the belief update judgments of LLMs diverge substantially from humans, which may complicate attempts that treat LLM responses as general-purpose proxies for human responses.

\section{Content Effects on LLM-Human Divergence (RQ2)}

Having established that LLM belief update judgments diverge from human labels across all models, we now ask what properties of the discussion content modulate the degree of this divergence. 
We examine three complementary dimensions, including textual features of the original post and reply (Section~\ref{sec::impact_of_textual_features}), the proposition type of the claim under debate (Section~\ref{sec::impact_of_proposition_types}), and the persuasion strategy employed in the reply (Section~\ref{sec::impact_of_persuasion_strategies}). 
These dimensions capture distinct aspects of the content: textual features operate at the surface level; proposition type captures what is being debated on the original post side; and persuasion strategy captures how the argument is constructed on the reply side.

\subsection{Impact of Textual Features} \label{sec::impact_of_textual_features}

\begin{figure}[ht]
    \centering
    \includegraphics[width=\linewidth]{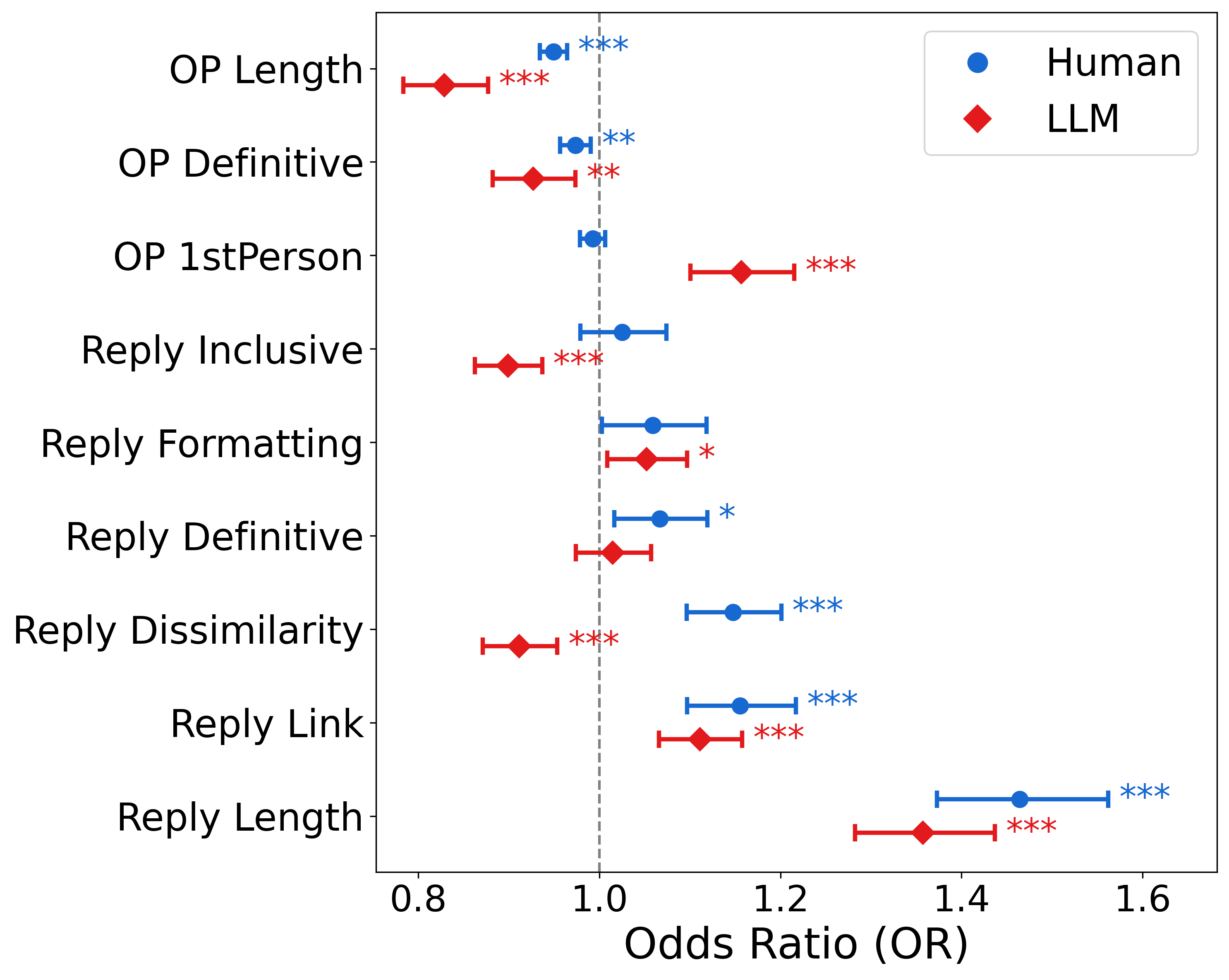}
    \caption{Logistic regression coefficients of belief update judgments on textual features. Significance is assessed using Benjamini--Hochberg FDR-corrected $q$-values, treating each regression as a separate family of nine tests ($^{*}q<0.05$, $^{**}q<0.01$, $^{***}q<0.001$).}
    \label{fig:regression_textual_features}
\end{figure}

To examine whether human and LLM belief updating rely on the same surface-level cues, we fit separate logistic regressions for humans and LLMs using the nine features described in Section~\ref{sec::extracting_textual_features}. 
The LLM regression pools judgments across all eight models with model fixed effects.
We report significance based on Benjamini--Hochberg FDR-corrected $q$-values, treating each regression as a separate family of nine tests.
Figure~\ref{fig:regression_textual_features} presents the odds ratio for each feature, sorted by human coefficient.
From the regression results, we make the following observations.

First, \textbf{four features influence human and LLM judgments in the same direction.}
\textit{Reply Length} is the strongest positive predictor for both (human $OR=1.4643$, LLM $OR=1.3574$, both $p<0.001$), suggesting that more elaborated arguments are regarded as more persuasive.
\textit{Reply Link} is also positively associated with persuasion for both (human $OR=1.1555$, LLM $OR=1.1109$, both $q<0.001$), indicating that citing external evidence increases credibility for humans and LLMs alike. 
On the original post side, \textit{OP Length} and \textit{OP Definitive} are both negatively associated with persuasion for both humans and LLMs, consistent with the intuition that more elaborated or more rigidly stated positions are harder to overturn. 
Nevertheless, the log-odds coefficient for \textit{OP Length} is over three times as strong in the LLM regression as in the human regression (human $OR=0.9493$, LLM $OR=0.8290$, both $q<0.001$), suggesting that LLMs are disproportionately resistant to persuasion when the original post is longer.

Second, \textbf{the remaining features diverge between humans and LLMs in sign or significance.}
The clearest case is \textit{Reply Dissimilarity}, where the two regressions yield opposite signs: for humans, replies that introduce novel content from their original post tend to be significantly more persuasive ($OR=1.1476$, $q<0.001$), whereas for LLMs the association reverses ($OR=0.9115$, $q<0.001$). 
This suggests that humans reward fresh perspectives that go beyond the original framing, while LLMs favor replies that stay topically close to the original post.

Three features show significant effects exclusively for LLMs.
\textit{Reply Formatting} is a borderline case: it shows a comparable positive point estimate for humans and LLMs ($OR=1.0591$ vs.\ $1.0520$), but survives FDR correction only for LLMs ($q=0.019$), with the human effect just missing the threshold ($q=0.050$).
This shared direction is consistent with the idea that structured presentation such as bullet points and numbered lists lowers processing load and makes an argument easier to follow and evaluate~\cite{o2008elaboration}.
\textit{Reply Inclusive} has no reliable association with human persuasion ($OR=1.0254$, $q=0.302$), but is negatively associated with LLM persuasion ($OR=0.8991$, $q<0.001$).
While inclusive language serves as a rapport-building strategy for human audiences, it may soften the perceived force of the argument from the LLM's perspective, reducing its persuasive impact.
\textit{OP 1stPerson} is similarly unrelated to human outcomes ($OR=0.9926$, $q=0.302$) yet strongly predicts LLM persuasion in the positive direction ($OR=1.1567$, $q<0.001$), suggesting that LLMs may interpret frequent first-person expression in the original post as a signal of subjective, personally held beliefs that are more amenable to revision.

Conversely, \textit{Reply Definitive} is positively associated with human persuasion ($OR=1.0668$, $q=0.013$) but not with LLM persuasion ($OR=1.014$, $q=0.480$), suggesting that assertive and confident language enhances perceived conviction for human readers while LLMs remain insensitive to this rhetorical signal.

We also re-estimate an OLS regression with a continuous, LLM-rated belief-change score (0--100) as the dependent variable in place of the binary outcome, and find the results remain consistent in direction and significance with our main findings (see Appendix~\ref{app::continuous_score}).

In sum, while humans and LLMs agree on the strongest and most robust drivers—longer, evidence-backed replies are more persuasive, and longer or more definitive original posts are harder to overturn—they diverge on a set of finer cues. Humans are distinctively swayed by the novelty of the reply and by confident delivery, whereas LLMs are comparatively more responsive to surface-level formatting and stylistic markers and reward replies that stay topically close to the original post.



\subsection{Impact of Proposition Types} \label{sec::impact_of_proposition_types}

To examine whether the type of claim under debate modulates LLM-human divergence, we classify each original post by its proposition type as described in Section~\ref{sec::proposition_type_annotation}.
Across all unique posts, value propositions account for the largest share (44.7\%), followed by fact propositions (28.0\%) and policy propositions (27.3\%). 

\begin{figure}[ht]
    \centering
    \includegraphics[width=\linewidth]{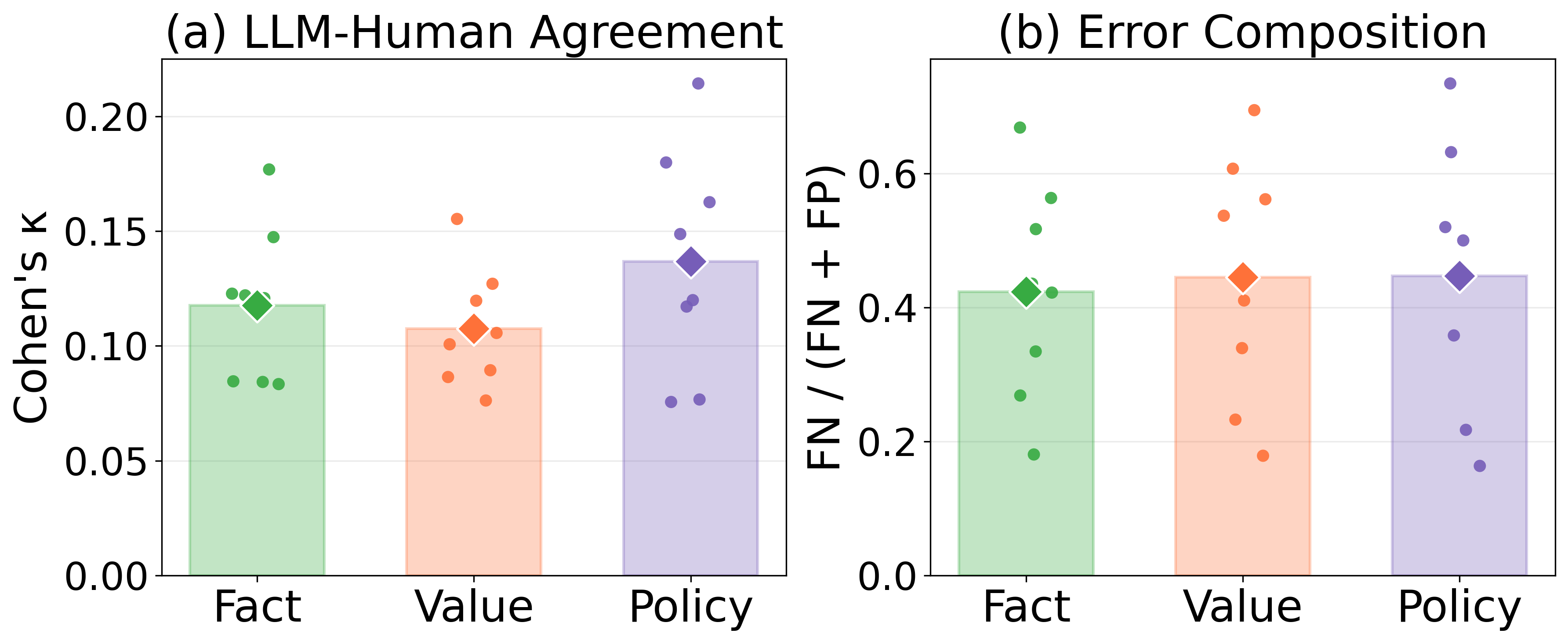}
    \caption{LLM judgment bias across proposition types.}
    \label{fig:proposition_type_impact}
\end{figure}

In Figure~\ref{fig:proposition_type_impact}, we break down LLM-human agreement and error composition by proposition type.
Across the three types, Cohen's $\kappa$ values remain in a similar range and show no systematic separation (panel a), and the error composition likewise shows no consistent shift toward resistance or receptiveness for any particular type (panel b). 
This suggests that LLM-human agreement does not meaningfully vary with proposition type. 
Thus, the type of claim under debate is not a meaningful moderator of LLM-human divergence.

\subsection{Impact of Persuasion Strategies} \label{sec::impact_of_persuasion_strategies}

\begin{figure}[ht]
    \centering
    \includegraphics[width=\linewidth]{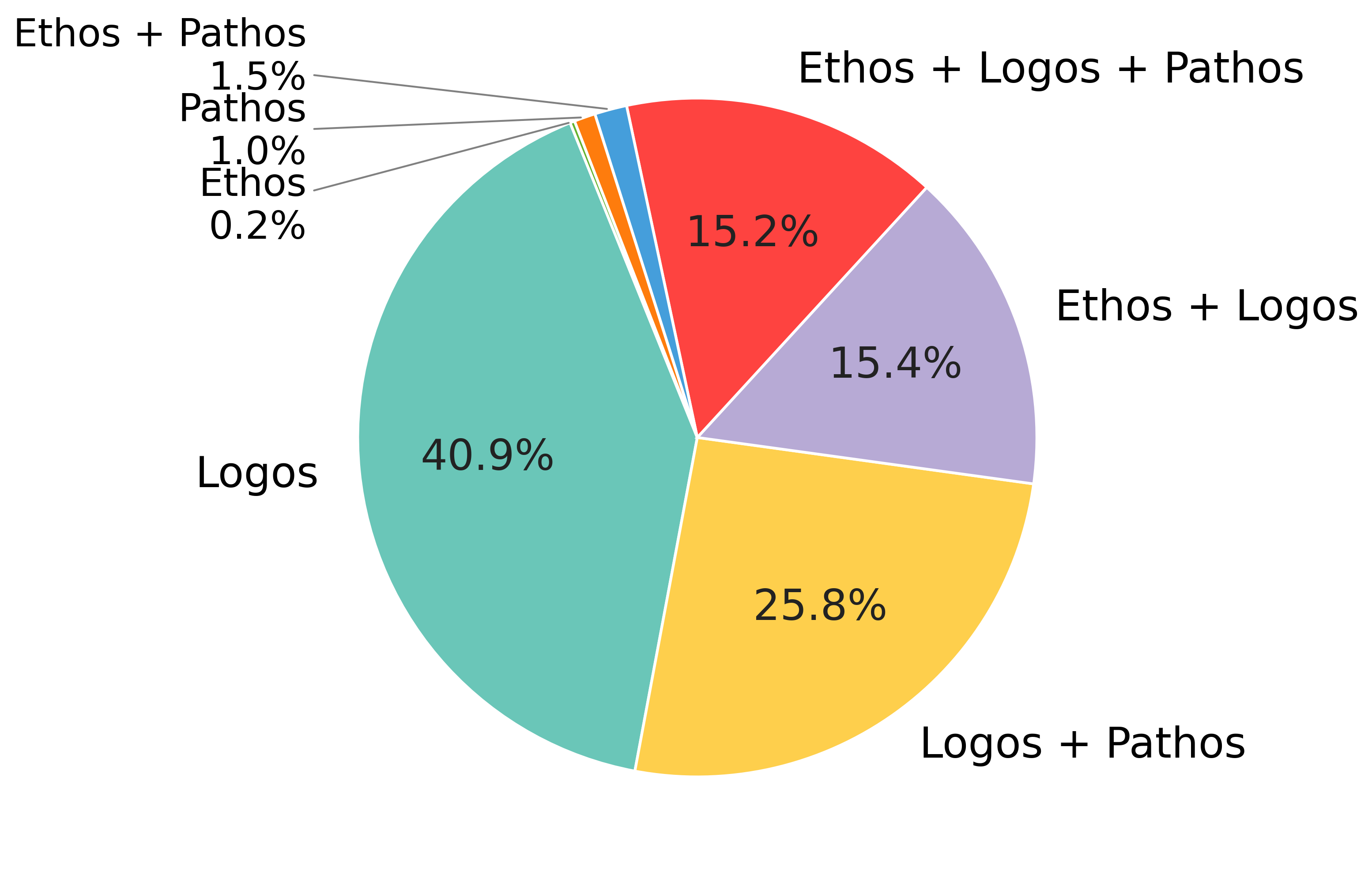}
    \caption{Distribution of persuasion strategies in the dataset.}
    \label{fig:distribution_strategy}
\end{figure}

To examine whether the persuasion strategy employed in a reply modulates LLM-human divergence, we first annotate each challenger reply according to the multi-label scheme described in Section~\ref{sec::persuasion_strategy_annotation}.
As shown in Figure~\ref{fig:distribution_strategy}, 42.2\% of replies rely on a single persuasion strategy, among which \textit{logos} dominates overwhelmingly (40.9\%) while \textit{pathos} alone (1.0\%) and \textit{ethos} alone (0.2\%) are rare. 
The remaining replies combine multiple strategies, with \textit{logos + pathos} (25.8\%), \textit{logos + ethos} (15.4\%), and all three (15.2\%) as the major combinations.

\begin{figure}[ht]
    \centering
    \includegraphics[width=\linewidth]{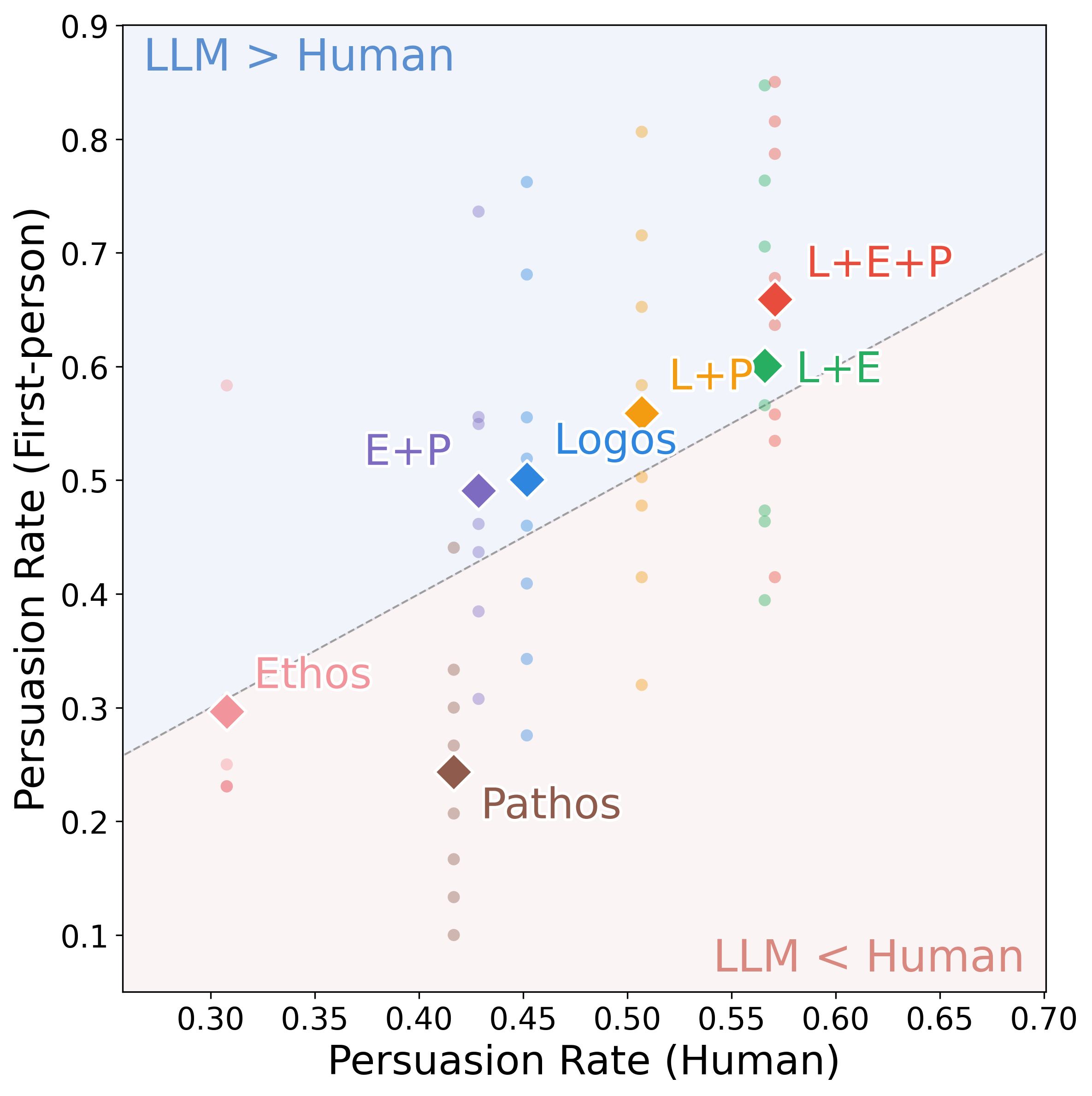}
    \caption{Effectiveness of persuasion strategies.}
    \label{fig:strategy_impact}
\end{figure}

Figure~\ref{fig:strategy_impact} plots the persuasion rate of each strategy combination for humans (x-axis) against LLMs (y-axis). 
\textbf{Humans and LLMs share two broad patterns}. 
First, combining multiple strategies yields higher persuasion rates than relying on a single one: the three-mode combination (L+E+P) achieves the highest rate on both sides (human 57.08\%, LLM 65.93\% on average).
Second, the \textit{ethos} alone achieves low persuasion rates for both humans and LLMs (30.77\% / 29.65\% on average), but adding it to other strategies consistently boosts effectiveness, suggesting that it functions as an amplifier rather than a standalone strategy.

Beyond these shared tendencies, \textbf{we also observe two divergences between humans and LLMs}. 
The largest gap is observed for the \textit{pathos} strategy, where humans achieve a moderate persuasion rate (41.67\%) but LLMs respond at a much lower rate (24.34\% on average), suggesting that purely emotional appeals carry little persuasive weight for LLMs.
Conversely, \textit{ethos}-containing combinations show the opposite pattern, with LLM persuasion rates consistently exceeding human rates (e.g., E+P: 42.86\% for humans vs.\ 49.08\% for LLMs on average).
This suggests that appeals grounded in authority or personal expertise carry disproportionate persuasive weight for LLMs.

Taken together, these results suggest that while LLMs and humans respond similarly to the structural complexity of arguments (more strategies, more persuasion), they diverge in their sensitivity to specific persuasive modes: LLMs underweight emotional engagement and overweight credibility signals relative to humans.

\section{Perspective-Induced Bias in Belief Update Judgments (RQ3)} \label{sec::RQ3}

Previous sections evaluate LLMs under the first-person condition, in which the model role-plays as the original poster.
A natural follow-up question is whether this perspective itself contributes to the observed bias. 
We address this by introducing a third-person condition, in which the model acts as an external observer predicting whether a reply would change the original poster's view (see Appendix~\ref{app::third_person_prompt} for the prompt template).

\begin{figure}[ht]
    \centering
    \includegraphics[width=\linewidth]{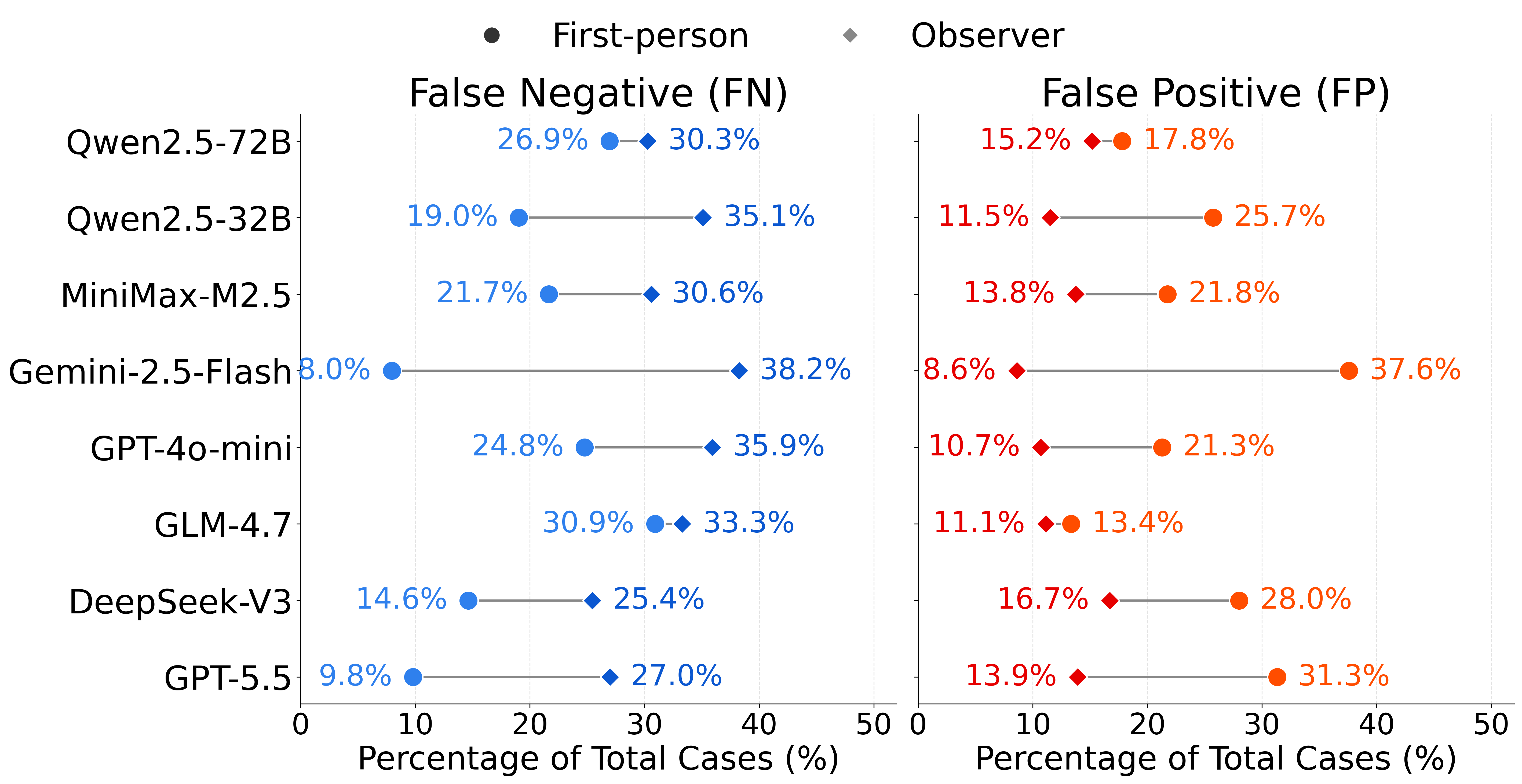}
    \caption{Comparing the agreement with human judgments across first-person and third-person LLM judgments.}
    \label{fig:perspective_comparison}
\end{figure}

First, \textbf{switching to the observer perspective increases resistance to persuasion.} 
As Figure~\ref{fig:perspective_comparison} shows, the internal composition of these errors shifts substantially. 
All eight models exhibit higher FN rates and lower FP rates under the observer condition than under the first-person condition. Whereas the first-person condition produces three distinct error profiles (Section 3), the observer condition collapses this variation: every model now displays a resistant profile in which FN far exceeds FP. 
The most striking case is \texttt{Gemini-2.5-Flash}, whose error composition nearly mirrors itself across conditions (first-person: FN = 8.0\%, FP = 37.6\%; observer: FN = 38.2\%, FP = 8.6\%).

This pattern suggests that first-person role-playing amplifies models' inclination to accommodate the argument they are confronted with, aligning with the sycophantic behavior widely documented in LLMs~\cite{sharma2024towards,ranaldi2023large}. 
When a model is instructed to evaluate a challenger's argument as the original poster, it faces the argument directly in the user turn and is more inclined to concede. 
When the same model predicts a third party's response, this pressure diminishes, and all models converge toward a conservative default of predicting that persuasion did not occur. 
Notably, this observer-side shift runs opposite to a well-established pattern in human communication. 
The third-person effect~\cite{davison1983third,sun2008understanding} holds that people perceive persuasive messages as more influential on others than on themselves, over-attributing susceptibility when reasoning about a third party.

\begin{figure}[ht]
    \centering
    \includegraphics[width=\linewidth]{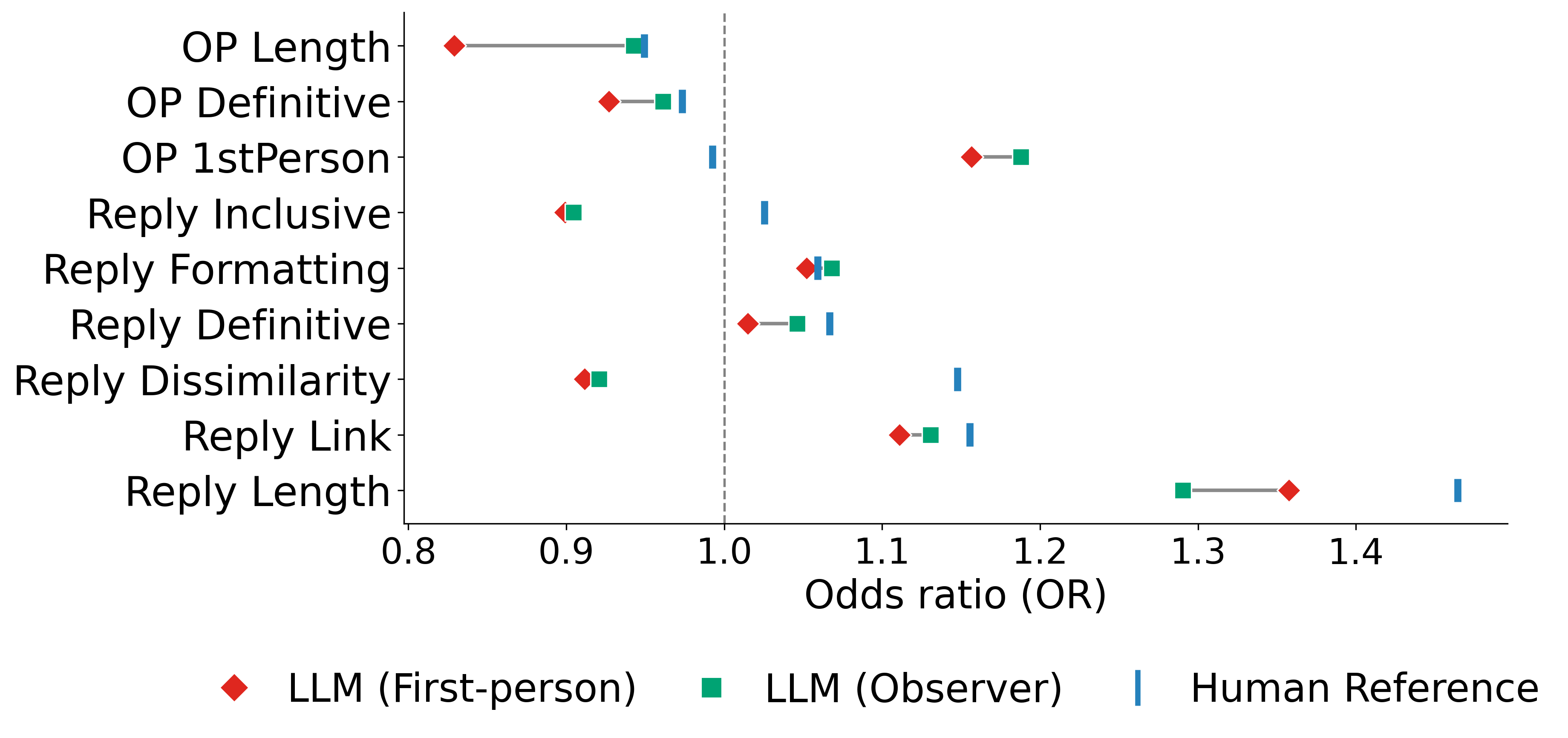}
    \caption{Change in logistic regression coefficients on textual features, when shifting from first-person to observer perspective.}
    \label{fig:regression_change}
\end{figure}

Second, \textbf{switching to the observer perspective partially realigns the textual feature-dependence structure with humans.}
As Figure~\ref{fig:regression_change} shows, for most features where the first-person LLM regression diverges from the human reference, the observer condition shifts the coefficient toward the human baseline. 
The largest corrections occur for \textit{OP Length}, where the negative association in the first-person LLM regression weakens substantially under the observer condition.
Two features move in the opposite direction, though: \textit{OP 1stPerson}, already a strong positive predictor only for LLMs, strengthens further.
For \textit{Reply Length}, the LLM coefficient decreases further from the first-person to the observer condition, widening the gap between LLMs and humans.
Nevertheless, the observer condition brings the majority of LLM feature coefficients closer to human values overall. 
One possible explanation is that first-person role-playing introduces a tension between internalizing the original poster's position and evaluating the incoming argument on its merits, and this tension distorts the feature weights that drive the model's judgments. 
The observer framing weakens this tension, allowing the model to evaluate arguments with less interference from its assigned role.

\begin{figure}[ht]
    \centering
    \includegraphics[width=\linewidth]{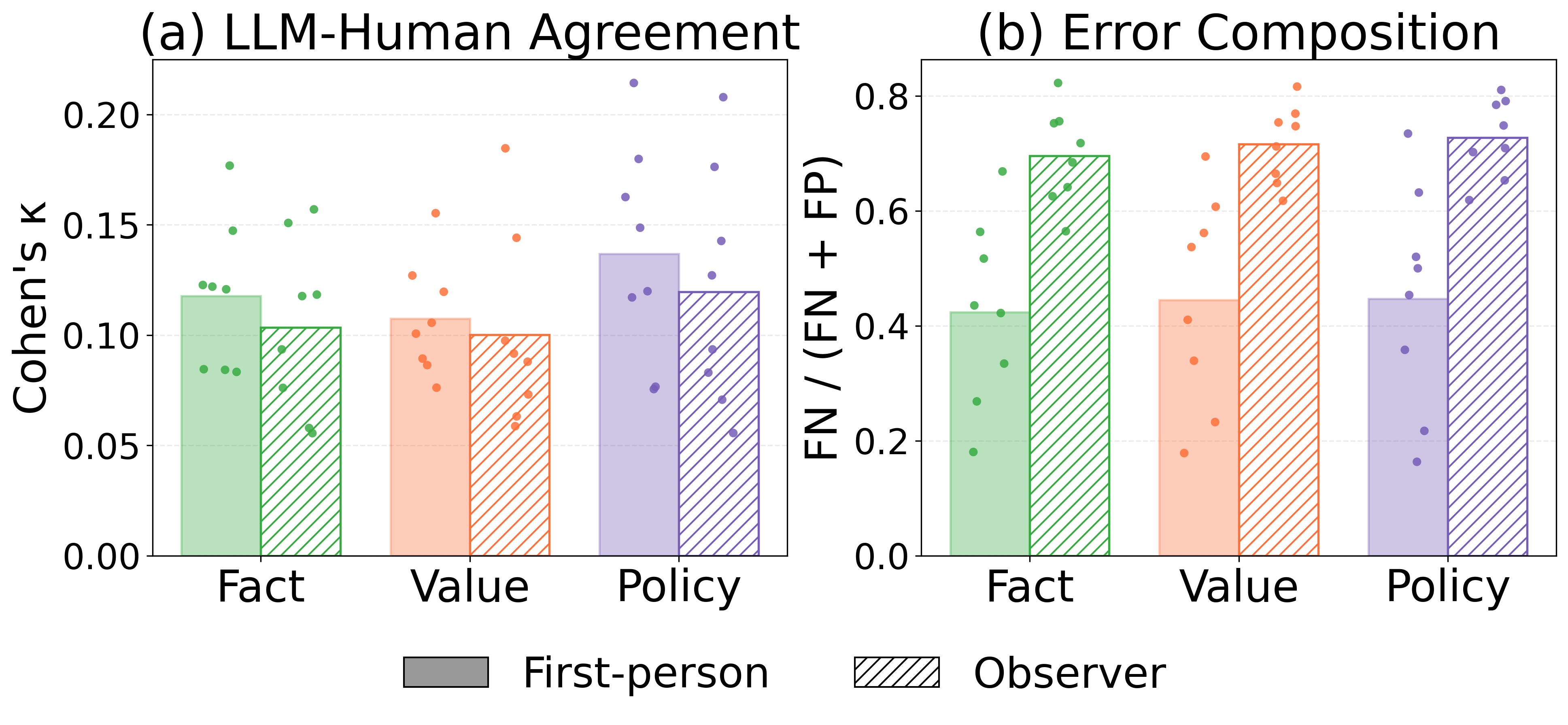}
    \caption{Comparison of first-person and observer perspectives by proposition type.}
    \label{fig:perspective_comparison_by_proposition_type}
\end{figure}

\begin{figure}[ht]
    \centering
    \includegraphics[width=\linewidth]{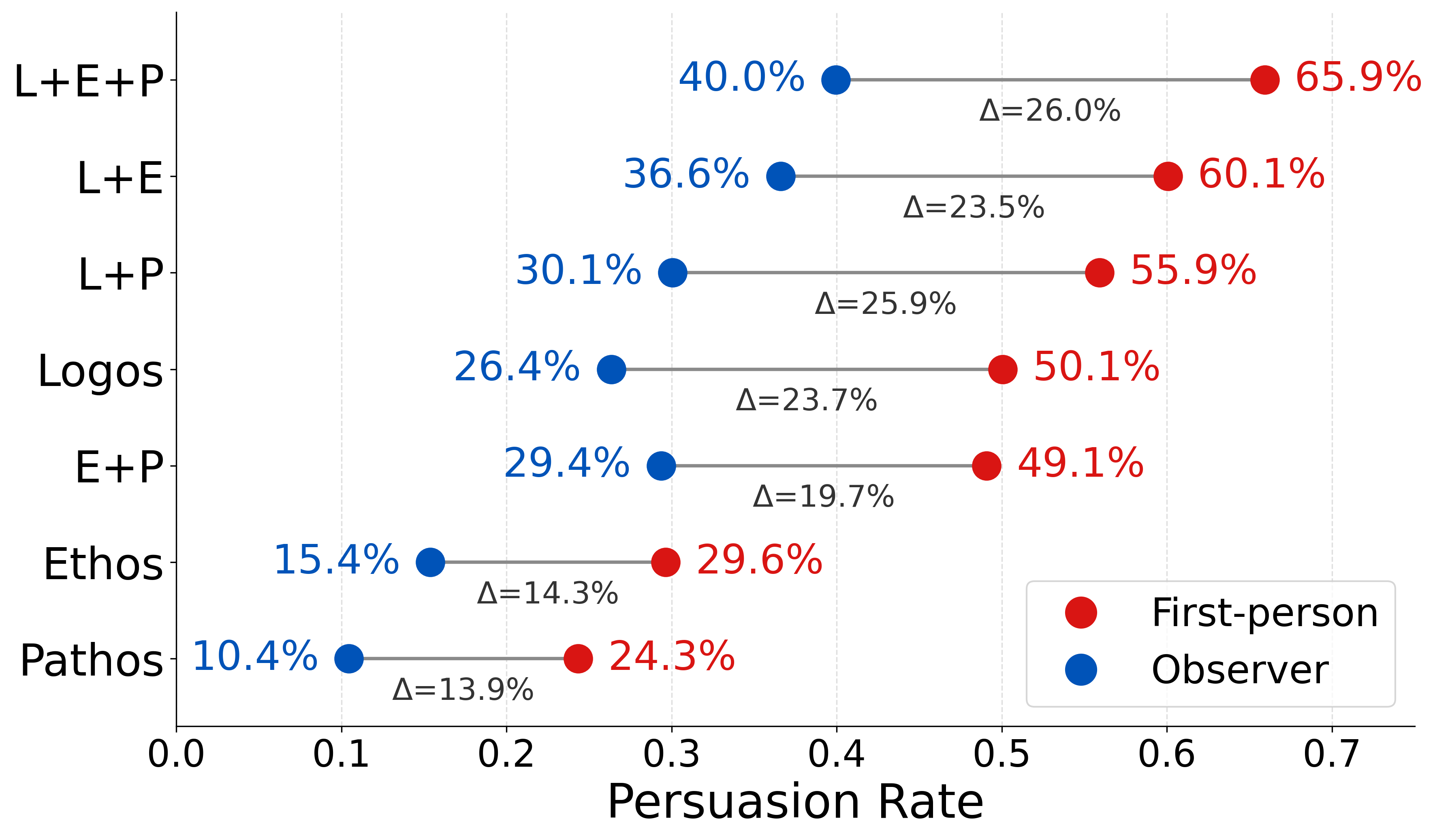}
    \caption{Comparison of first-person and observer perspectives by persuasion strategy.}
    \label{fig:strategy_effectiveness_first_person_observer}
\end{figure}

Third, \textbf{the perspective shift has a differential impact on persuasion rates across strategy types, while proposition type continues to show no effect.}
As Figure~\ref{fig:perspective_comparison_by_proposition_type} shows, the type of proposition under debate remains a non-moderator of LLM–human divergence under the observer condition.
At the level of persuasion strategy, Figure~\ref{fig:strategy_effectiveness_first_person_observer} shows that all strategy combinations yield lower LLM persuasion rates under the observer condition, consistent with the overall increase in resistance. 
The rank ordering of strategies is largely preserved across both perspectives, with pathos being the least effective strategy and the tri-strategy combination being the most effective.
However, the absolute decline varies with the type of strategy employed. 
Strategies containing logos show larger declines in effectiveness ($\Delta \approx 23.5$--$26.0~pp$), while strategies without logos show smaller declines ($\Delta \approx 13.9$--$19.7~pp$), even at comparable baseline rates (e.g., \textit{E+P} at 49.1\% vs. \textit{Logos} at 50.1\%). 
This indicates that the perspective effect is not a uniform threshold shift, but interacts with the persuasive content of the argument.

In sum, the evaluation perspective modulates the form of LLM–human divergence in belief update judgments. 
The observer condition suppresses the accommodative tendencies present in first-person role-playing, partially realigns feature-level sensitivities with humans, yet affects different persuasion strategies to different degrees. 
These results suggest that perspective is an active source of bias in LLM belief updating rather than a neutral design choice.

\section{Related Work}

\paragraph{LLMs as proxies for humans.}
A growing body of work employs LLMs as stand-ins for human participants, from generative agents and simulated social interaction~\cite{park2023generative,gao2024large} to opinion and survey simulation~\cite{argyle2023out}, online discourse~\cite{chuang2024simulating}, elections~\cite{zhang2024electionsim}, and collective decision-making~\cite{jarrett2025language}. 
Yet the fidelity of such proxies is contested: while LLMs can reproduce human response distributions at the demographic level~\cite{argyle2023out}, they misalign for specific subgroups and topics~\cite{santurkar2023whose}, prompting calls for more careful evaluation of their values and opinions~\cite{rottger2024political}. 
However, this literature assesses fidelity mainly on static opinion distributions and survey responses, and it is less clear whether the same fidelity extends to belief updating, that is, to matching humans in how a view is revised or maintained in response to a specific persuasive argument. 
We address this by benchmarking LLM belief-update judgments against participant-verified human labels on real single-round persuasion data.

\paragraph{LLM persuasion and opinion change.}
LLMs are increasingly studied as both producers and evaluators of persuasion. 
Controlled experiments show that LLM-generated arguments can be as persuasive as, or more persuasive than, human ones, especially when personalized to the audience~\cite{salvi2025conversational,breum2024persuasive,matz2024potential}. 
In multi-agent and debate settings, LLMs exhibit systematic tendencies such as accuracy bias driving group consensus~\cite{chuang2024simulating} and inherent social biases overriding assigned ideological stances~\cite{taubenfeld2024systematic}. 
Most of this work, however, measures whether LLMs \emph{generate} persuasive text or how they behave in open-ended debate, rather than benchmarking their belief-update \emph{judgments} against human ground truth on individual arguments. 
We instead build on the CMV corpus~\cite{tan2016winning}, whose participant-verified deltas provide such ground truth, and ask which content-level properties modulate human--LLM divergence.

Two works are closest to our setting, both using the same CMV corpus that we do. 
\citet{donmez2025understand} study LLMs as persuaders: they generate counter-arguments on CMV and compare the social dimensions (e.g., trust, similarity) expressed in LLM-generated vs.\ human-written arguments, finding that LLMs convey richer pragmatic cues. 
\citet{freeborn2026change} use LLMs as analytical tools: they forecast persuasion outcomes in CMV discussions and code rhetorical strategies to predict delta awards. 
In both cases, the object of study is the argument itself, not whether an LLM, as the recipient, updates its belief the way a human would. 
Our work instead studies LLMs as persuadees: we present the same arguments that humans evaluated and benchmark LLM belief-update judgments directly against human-verified outcomes. 
Rather than asking what makes LLM-generated arguments persuasive, we ask whether LLMs respond to persuasive arguments the way humans do, and diagnose the divergence along three complementary dimensions.


\paragraph{Perspective framing, sycophancy, and self--other asymmetries.}
Several lines of work bear on how perspective and role shape LLM judgments. 
Assigning different roles to LLMs, such as actor versus observer, has been shown to induce systematic attribution biases~\citep{li2026taming}, and a growing literature probes the theory-of-mind and perspective-taking abilities of LLMs, finding them non-trivial but imperfect~\citep{kosinski2024evaluating,strachan2024testing}. 
A related concern is \emph{sycophancy}, the tendency of models to favor agreement with the interlocutor over accuracy.
It has been traced to preference-optimization training and only partially mitigated by subsequent interventions~\citep{sharma2024towards,ranaldi2023large,hao2026reinforcement}. 
On the human side, the classical \emph{third-person effect} holds that people perceive persuasive messages as more influential on others than on themselves~\citep{davison1983third,sun2008understanding}, offering a psychological baseline against which the perspective behavior of LLMs can be compared. 
Yet how these role, sycophancy, and self--other effects manifest specifically in belief-update judgments, and whether they align with or depart from the human pattern, remains largely unexamined. 
We address this gap by manipulating first-person versus observer framing over the same arguments and comparing the resulting LLM perspective effects against this human baseline.

\section{Conclusion}

In this study, we systematically compare LLM belief update judgments against human-verified persuasion outcomes.
All tested models achieve only slight agreement with human labels. 
The overall rate of divergence is similar across models, but its internal composition differs markedly.
Diagnosing what drives this divergence, we find that it is insensitive to the type of claim under debate but systematically shaped by how arguments are constructed. 
LLMs and humans rely on qualitatively different cues when evaluating persuasive force, with LLMs favoring topical overlap and credibility signals while discounting novelty and emotional engagement. 
Introducing a third-person observer perspective shifts all models toward greater resistance to persuasion, but the effect is uneven across persuasion strategies and textual features.
These patterns point to a structural mismatch between LLM and human belief updating.
Such a mismatch persists even in the strongest model tested, and thus whether further scaling resolves it remains an empirical question for future analysis.
Moreover, future work may extend this evaluation to multi-turn persuasion trajectories and examine whether these divergence patterns generalize across languages and demographic contexts.

\newpage
\section*{Limitations}

Our study has several limitations.
First, in the \textit{CMV} corpus, user population likely skews toward young, English-speaking, politically engaged Reddit users~\cite{pew2016reddit}, which may limit the generalizability of our findings to other demographic contexts. 
That said, it remains one of the few publicly available resources that provides the three-part structure essential for this study: an initial belief, a counterargument, and a participant-verified persuasion outcome. 
The ground-truth labels themselves carry two further limitations: they reflect only whether the original poster acknowledged being persuaded in direct response to a single root reply, introducing asymmetric noise into the negative labels when persuasion occurred through subsequent exchanges; and they compress a spectrum of belief change into a binary outcome, collapsing partial concessions and complete shifts into a single positive label. 
We adopt the single-turn design to isolate the persuasive effect of an individual argument; extending the evaluation to multi-turn persuasion trajectories is a natural direction for future work.


Second, our experiments evaluate all models under default configurations without injecting persona information such as the original poster's values, background, or prior discussion history. 
Since this contextual information is unavailable in the dataset and varies across individuals, the default configuration serves as a principled baseline rather than a faithful simulation of any specific human respondent. 
A promising direction is to augment the dataset with richer persona attributes, and evaluate LLMs' capability of simulating these personalized views.

Third, although the examined LLMs span multiple model families and language ecosystems, the rapid pace of model development means that specific bias magnitudes may shift in future generations. 
Nevertheless, the structural patterns we identify likely reflect broader properties of current pretraining and alignment paradigms rather than idiosyncrasies of individual checkpoints.

\section*{Ethical Considerations}

\paragraph{Data and Privacy.}
All data used in this study are drawn from the publicly available ChangeMyView corpus.
Reddit usernames and other personally identifiable information are not used in any of our analyses. 
Since our experiments involve only LLM inference on existing public data, no human participants were recruited and no IRB approval was required.

\paragraph{Sensitive Content.}
CMV discussions occasionally address sensitive or controversial social, political, and moral topics. 
Our annotation prompt explicitly instructs the annotation model to treat these texts as research material and classify them according to the defined schema, without making editorial judgments about their content.

\paragraph{Potential for Misinterpretation.}
Our findings reveal systematic divergences between LLM and human belief updating behavior. We caution against interpreting these results as evidence that LLMs are defective reasoners; rather, they highlight the risk of treating LLM judgments as faithful proxies for human responses in simulation and decision-support applications without careful validation.


\bibliography{main}

\newpage
\appendix

\section*{Appendix}

\section{Prompt Templates}  \label{app:prompt_templates}

\subsection{First-person belief update judgment} \label{app::first_person_prompt}

Figure~\ref{fig:prompt_first_person} shows the prompt template for first-person belief update judgment.
The system prompt assigns the model the role of the original poster and presents the post title and body, so that the model first internalizes the poster's position. The user prompt then presents the challenger's reply, instructs the model to judge whether the reply has changed its view, and provides explicit guidelines for awarding a delta: a delta should be awarded when the reply has meaningfully changed the model's position, and does not require complete abandonment of the original view. The model outputs a JSON object containing a binary delta decision and a free-text justification.

\begin{figure}[ht] 
\centering
\small
\textbf{System prompt}
\begin{lstlisting}[style=prompt]
"""You are the Original Poster (OP) who wrote the following post on an online forum.
   Title: {title}
   Body: {text}
"""
\end{lstlisting}
\textbf{User prompt (template)}
\begin{lstlisting}[style=prompt]
"""A challenger has replied to your post with the following argument:

Challenger's Reply:
{challenger_text}

As the OP, evaluate whether this reply has changed your view.

AWARDING A DELTA:
Award a delta ("delta_awarded": true) if the challenger's reply has actually changed your mind on this issue. This means your original claim, as stated in your post, no longer fully represents what you believe. The change can be a total shift or a substantial adjustment of your original stance.

Output your response strictly as a JSON object.

JSON RULES:
1. Use DOUBLE QUOTES for all keys and string values.
2. Escape any internal double quotes with a backslash.
3. Use \\n instead of actual line breaks inside string values.

Format:
{{
"delta_awarded": <true or false>,
"justification": "<Explain why the reply did or did not change your view.>"
}}
"""

\end{lstlisting}
\caption{Prompt template for first-person evaluation of persuasion.}
\label{fig:prompt_first_person}
\end{figure}

\subsection{Third-person (observer) belief update judgment} \label{app::third_person_prompt}

Figure~\ref{fig:prompt_third_person} shows the prompt template for third-person (observer) belief update judgment.
The template mirrors the first-person version (Figure~\ref{fig:prompt_first_person}) in structure and wording, with two key modifications: the system prompt re-frames the model as an impartial external observer rather than the original poster, and the user prompt changes the task from evaluating whether "your view" has changed to predicting whether the reply would have changed the poster's view. This parallel design ensures that any behavioral difference between the two conditions can be attributed to the perspective shift itself rather than to incidental prompt variation.

\begin{figure}[!htbp]
\centering
\small
\textbf{System prompt}
\begin{lstlisting}[style=prompt]
"""You are an impartial external observer reading an exchange on an online forum. The Original Poster (OP) wrote the following post.
Title: {title}
Body: {text}
"""
\end{lstlisting}
\textbf{User prompt (template)}
\begin{lstlisting}[style=prompt]
"""A challenger has replied to the OP's post with the following argument:

Challenger's Reply:
{challenger_text}

As an external observer, predict whether this reply has changed the OP's view.

AWARDING A DELTA:
Predict whether a delta is awarded. A delta should be awarded if the challenger's reply would actually change the OP's mind on this issue. This means the OP's original claim, as stated in their post, would no longer fully represent what they believe. The change can be a total shift or a substantial adjustment of the OP's original stance.

Output your response strictly as a JSON object.

JSON RULES:
1. Use DOUBLE QUOTES for all keys and string values.
2. Escape any internal double quotes with a backslash.
3. Use \\n instead of actual line breaks inside string values.

Format:
{{
"delta_awarded": <true or false>,
"justification": "<Explain why you predict the reply would or would not change the OP's view.>"
}}
"""

\end{lstlisting}
\caption{Prompt template for third-person evaluation of persuasion.}
\label{fig:prompt_third_person}
\end{figure}

\subsection{Annotation of proposition type} \label{app::proposition_type_annotation_prompt}

Figure~\ref{fig:prompt_proposition_annotation} shows the prompt template for annotating the proposition type of an original post.
The system prompt defines three proposition types following \citet{freeley2009argumentation}, each illustrated with two verified examples.
Because many posts contain elements of more than one type, a classification rule instructs the model to focus on the core claim the poster is ultimately defending rather than the supporting arguments. 
The user prompt presents the original post title and body, and the model outputs a single-label JSON object assigning the post to one of the three categories.

\begin{figure}[!htbp]
\centering
\small
\textbf{System prompt}
\begin{lstlisting}[style=prompt]
"""You are an expert annotator for argument topic classification.
Your task is to read an original post from an online debate forum and classify the core claim of the post into one of three proposition types.

PROPOSITION OF FACT: The core claim is about how the world IS. The disagreement could in principle be resolved through evidence, data, or logical demonstration. The poster is asserting that something is true or false, possible or impossible, or making a causal or predictive claim.
Examples:
- "GMO foods are harmful to human health."
- "Self-driving cars are safer than human drivers."

PROPOSITION OF VALUE: The core claim is a judgment about what is good, bad, right, wrong, important, or worthless. The disagreement is fundamentally about values, morals, or evaluative criteria, not purely about facts.
Examples:
- "The death penalty is morally unacceptable."
- "Young people who smoke despite knowing the risks are being foolish."

PROPOSITION OF POLICY: The core claim is about what should or should not be done. The disagreement is about a proposed course of action, law, rule, or behavioral norm.
Examples:
- "Marijuana should be legalized."
- "Governments should ban all surveillance programs."

CLASSIFICATION RULE: Many posts contain elements of more than one type. Classify based on the core claim the poster is defending, not the supporting arguments. Ask yourself: what is the poster ultimately trying to convince others of? If it is that something IS the case, choose FACT. If it is that something is GOOD or BAD, choose VALUE. If it is that something SHOULD BE DONE, choose POLICY."""
\end{lstlisting}
\textbf{User prompt (template)}
\begin{lstlisting}[style=prompt]
"""Original Post:
Title: {title}
Body: {text}

Classify the core claim of this post into one of three types.
Output strictly as a JSON object:
{{
"proposition_type": "<fact | value | policy>"
}}"""
\end{lstlisting}
\caption{Prompt template for annotating the proposition type of a post.}
\label{fig:prompt_proposition_annotation}
\end{figure}

\subsection{Annotation of persuasion strategy} \label{app::strategy_annotation_prompt}

Figure~\ref{fig:prompt_strategy_annotation} shows the prompt template for annotating the persuasion strategy (strategies) adopted in a challenger reply.
The system prompt defines three classical modes of persuasion (\textit{logos}, \textit{pathos}, and \textit{ethos}) with examples adapted from \citet{hidey2017analyzing}, and clarifies that a single reply may combine multiple strategies. 
Because CMV discussions occasionally involve sensitive topics that may trigger the content control of the annotation model, the system prompt begins with an explicit instruction to treat all texts as research material and annotate according to the schema without refusing based on topic sensitivity. 
The user prompt presents both the original post and the challenger's reply, and the model outputs a multi-label binary JSON object indicating the presence or absence of each mode independently.

\begin{figure}[!htbp]
\centering
\small
\textbf{System prompt}
\begin{lstlisting}[style=prompt]
"""
This is an academic research annotation task. The texts you will read are from a publicly available research dataset (Reddit ChangeMyView). Some posts may discuss sensitive topics. Your role is strictly to classify the text according to the given schema. Do not refuse to annotate based on topic sensitivity.

You are an expert annotator for persuasion strategy classification.
Your task is to read a challenger's reply to an original post on an online forum, and determine which persuasion modes the reply employs. A single reply may use one or more of the following modes.

Definitions and examples (adapted from Hidey et al., 2017):

LOGOS: Appeals to reason through logical argument, factual evidence, relevant examples, statistics, or causal reasoning.
Examples:
- "Eating healthy makes you live longer. The oldest man in the US followed a strictly fat-free diet."
- "He will probably win the election. He is the favorite according to the polls."

PATHOS: Appeals to the audience's emotions, empathy, or sense of identification. This includes evoking fear, sympathy, moral concern, or describing scenarios the audience can personally relate to.
Examples:
- "Doctors should stop prescribing antibiotics at a large scale. The spread of antibiotics will be a threat for the next generation."
- "You should put comfy furniture into your place. The feeling of being home is unforgettable."

ETHOS: Appeals to credibility established through personal experience, professional expertise, or reference to authoritative sources.
Examples:
- "I assure you the consequences of fracking are terrible. I have been living next to a pipeline since I was a child."
- "I trust his predictions about climate change. He is a Nobel Prize winner."

Note: A reply may combine multiple modes. For example, a reply that cites statistics (logos) while also sharing a personal story to evoke empathy (pathos) should be labeled as both logos and pathos."""

\end{lstlisting}
\end{figure}

\begin{figure}[!htbp] 
\centering
\small
\textbf{User prompt (template)}
\begin{lstlisting}[style=prompt]
"""Original Post:
Title: {title}
Body: {text}

Challenger's Reply:
{challenger_text}

For each persuasion mode, determine whether the challenger's reply employs it.
Output strictly as a JSON object:
{{
"logos": <true or false>,
"pathos": <true or false>,
"ethos": <true or false>
}}"""
\end{lstlisting}
\caption{Prompt template for annotating the persuasion strategies adopted by a reply post.}
\label{fig:prompt_strategy_annotation}
\end{figure}

\clearpage

\section{Judgment Stability across Repeated Runs} \label{app::stability_across_runs}

To assess robustness to sampling variation, we repeat the full first-person judgment experiment three times with different random seeds for all eight models. 
As Table~\ref{tab:stability_across_runs} shows, seven of the eight models exhibit high cross-run stability (Fleiss' $\kappa$ ranging from 0.739 to 0.900, pairwise agreement $\geq$ 86\%), indicating that stochastic variation has negligible impact on their reported results. 
The one exception is \texttt{MiniMax-M2.5}, whose judgments are only moderately stable across runs (Fleiss' $\kappa$ = 0.438, pairwise agreement $\geq$ 71\%, three-run exact match 58.5\%); its results should therefore be interpreted with additional caution.
  

\begin{table}[ht]
\centering
\caption{Judgment stability across three runs with different random seeds.}
\label{tab:stability_across_runs}
\small
\setlength{\tabcolsep}{4pt}
\begin{tabular*}{\linewidth}{@{\extracolsep{\fill}}lccc@{}}
\toprule
Model & Fleiss' $\kappa$ & \makecell{Pairwise \\ Agreement} & \makecell{Three-run \\ Exact Match} \\
\midrule
\texttt{GPT-4o-mini} & 0.900 & $\geq$ 94.2\% & 92.5\% \\
\texttt{Qwen2.5-72B-Instruct} & 0.890 & $\geq$ 94.6\% & 92.0\% \\
\texttt{DeepSeek-V3} & 0.888 & $\geq$ 94.5\% & 92.2\% \\
\texttt{Gemini-2.5-Flash} & 0.888 & $\geq$ 96.2\% & 94.5\% \\
\texttt{Qwen2.5-32B-Instruct} & 0.875 & $\geq$ 93.5\% & 90.8\% \\
\texttt{GPT-5.5} & 0.835 & $\geq$ 92.8\% & 89.8\% \\
\texttt{GLM-4.7} & 0.739 & $\geq$ 86.7\% & 81.5\% \\
\texttt{MiniMax-M2.5} & 0.438 & $\geq$ 71.0\% & 58.5\% \\
\bottomrule
\end{tabular*}
\end{table}

\section{Validation of LLM Annotation} \label{app::annotation_validation}

To validate the LLM annotations of both proposition types and persuasion strategies, we conduct two additional checks. 
First, we use \texttt{GLM-5.2} to re-annotate the full corpus. 
For proposition type, the two models achieve Cohen's $\kappa$ = 0.751 ("substantial"). 
For persuasion strategy, micro-averaged $\kappa$ = 0.817 ("almost perfect"). 
Second, two human annotators from the author team independently labeled a random sample of 100 items for each dimension. 
For proposition type, inter-annotator agreement is $\kappa$ = 0.450; \texttt{GPT-5.1}'s agreement with each annotator ($\kappa$ = 0.454 and 0.683) meets or exceeds this human baseline. 
For persuasion strategy, inter-annotator agreement is micro $\kappa$ = 0.660; \texttt{GPT-5.1}'s agreement with each annotator (micro $\kappa$ = 0.570 and 0.735) brackets this baseline. 
These results indicate that \texttt{GPT-5.1} performs within the range of human annotator variation. 

\section{Consistency across Models} \label{app::consistency_across_models}

\begin{figure}[ht]
    \centering
    \includegraphics[width=\linewidth]{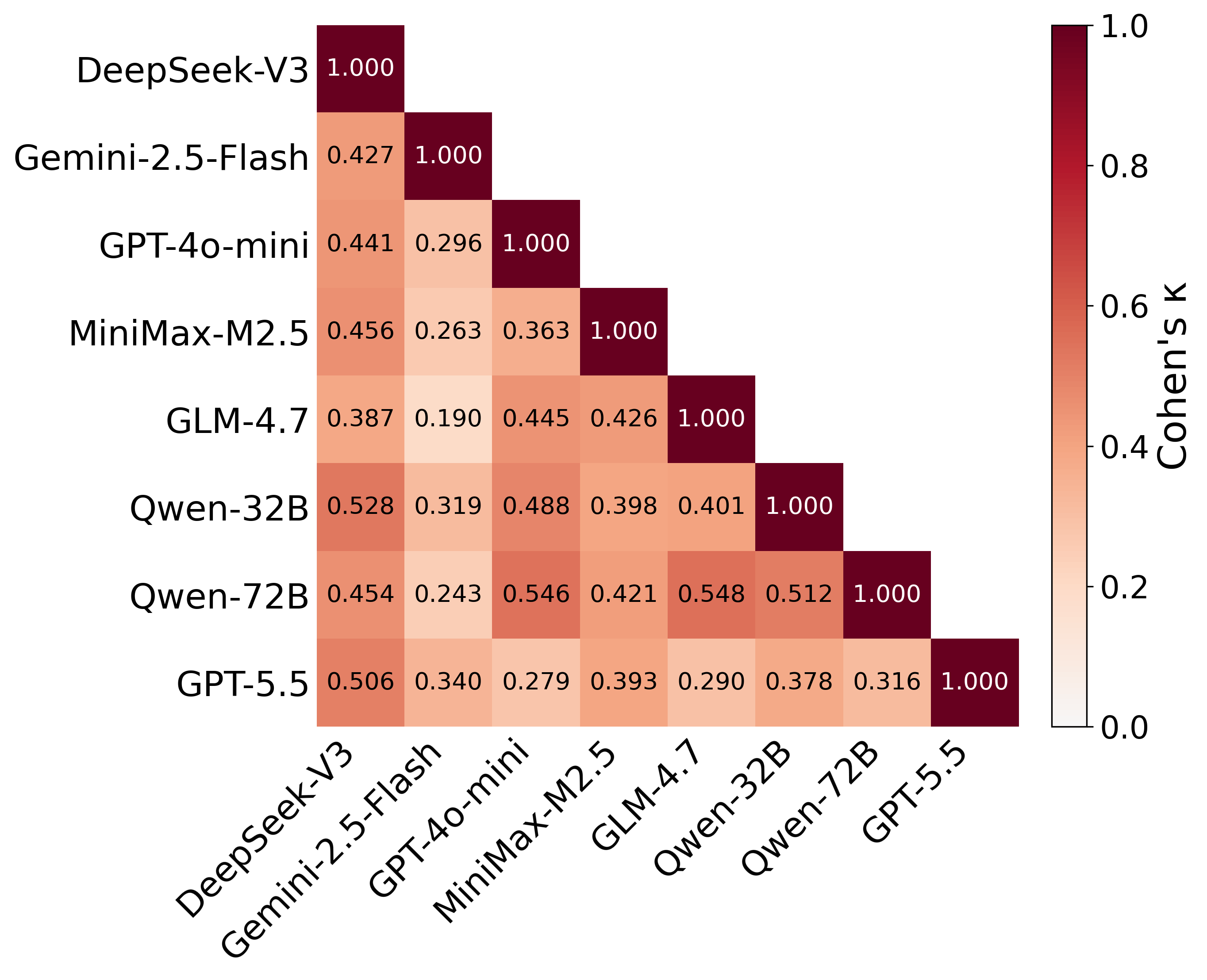}
    \caption{Belief update judgment consistency across different LLMs.}
    \label{fig:model_kappa}
\end{figure}

Figure~\ref{fig:model_kappa} presents pairwise Cohen's $\kappa$ values among the eight LLMs under the first-person condition. 
Inter-model agreement ranges from 0.190 (\texttt{GLM-4.7} vs. \texttt{Gemini-2.5-Flash}) to 0.548 (\texttt{Qwen-72B} vs. \texttt{GLM-4.7}), substantially exceeding the model-human agreement reported in Section~\ref{sec::RQ1}.

\section{Additional Textual Features}

Beyond the nine features analyzed in Section~\ref{sec::impact_of_textual_features}, Table~\ref{tab:additional_textual_features} reports the coefficient direction and p-value for the full set of textual features computed by \citet{tan2016winning}, from separate logistic regressions for humans and LLMs (pooled across models with model fixed effects). 
The majority of these features show no significant association with persuasion outcomes for either humans or LLMs. 
A small number reach significance exclusively for LLMs, including type-token ratio, paragraph count, and the fraction of OP vocabulary shared with the reply, suggesting that LLMs are sensitive to certain distributional properties of the text that carry little predictive value for human judgment.

\begin{table*}[t]
\centering
\small 
\caption{Comparison between humans and LLMs along the impact of additional textual features. All features are z-standardized before fitting. We report raw $p$-values and Benjamini--Hochberg FDR-corrected $q$-values, treating each regression as a separate family of 40 tests.}
\label{tab:additional_textual_features}
\begin{tabular}{l ccc ccc}
\toprule
\multirow{2}{*}{Feature} & \multicolumn{3}{c}{Human} & \multicolumn{3}{c}{LLM (First-person)} \\
\cmidrule(lr){2-4}\cmidrule(lr){5-7}
 & Trend & $p$ & $q$ & Trend & $p$ & $q$ \\
\midrule
num\_words & $\downarrow$ & 0.125 & 0.417 & $\downarrow$ & 0.671 & 0.839 \\
num\_definite\_articles & $\uparrow$ & 0.385 & 0.642 & $\downarrow$ & 0.030 & 0.134 \\
num\_indefinite\_articles & $\uparrow$ & 0.355 & 0.642 & $\uparrow$ & 0.293 & 0.589 \\
num\_positive\_words & $\uparrow$ & 0.292 & 0.584 & $\uparrow$ & 0.300 & 0.589 \\
num\_2nd\_person\_pronouns & $\uparrow$ & 0.068 & 0.277 & $\downarrow$ & 0.286 & 0.589 \\
num\_links & $\uparrow$ & 0.911 & 0.934 & $\downarrow$ & 0.445 & 0.741 \\
num\_negative\_words & $\uparrow$ & 0.382 & 0.642 & $\downarrow$ & 0.871 & 0.917 \\
num\_hedges & $\downarrow$ & 0.862 & 0.907 & $\uparrow$ & 0.165 & 0.414 \\
num\_1st\_person\_pronouns & $\uparrow$ & 0.802 & 0.907 & $\uparrow$ & 0.339 & 0.589 \\
num\_1st\_person\_plural\_pronouns & $\uparrow$ & 0.436 & 0.671 & $\downarrow$ & $<$.001 & $<$.001 \\
num\_dotcom\_links & $\downarrow$ & 0.209 & 0.521 & $\uparrow$ & 0.491 & 0.758 \\
frac\_links & $\uparrow$ & 0.163 & 0.469 & $\uparrow$ & 0.135 & 0.361 \\
frac\_dotcom\_links & $\uparrow$ & 0.164 & 0.469 & $\uparrow$ & 0.339 & 0.589 \\
num\_examples & $\uparrow$ & 0.235 & 0.523 & $\uparrow$ & 0.135 & 0.361 \\
frac\_definite\_articles & $\uparrow$ & 0.548 & 0.756 & $\uparrow$ & 0.801 & 0.915 \\
num\_question\_marks & $\downarrow$ & 0.973 & 0.973 & $\downarrow$ & 0.071 & 0.236 \\
num\_pdf\_links & $\uparrow$ & 0.724 & 0.860 & $\uparrow$ & 0.041 & 0.162 \\
frac\_positive\_words & $\downarrow$ & 0.015 & 0.151 & $\downarrow$ & 0.062 & 0.227 \\
arousal & $\downarrow$ & 0.430 & 0.671 & $\downarrow$ & 0.990 & 0.994 \\
valence & $\uparrow$ & 0.069 & 0.277 & $\downarrow$ & 0.598 & 0.825 \\
word\_entropy & $\downarrow$ & 0.855 & 0.907 & $\uparrow$ & $<$.001 & 0.003 \\
num\_sentences & $\downarrow$ & 0.470 & 0.696 & $\downarrow$ & 0.570 & 0.814 \\
type\_token\_ratio & $\downarrow$ & 0.056 & 0.277 & $\downarrow$ & $<$.001 & $<$.001 \\
num\_paragraphs & $\downarrow$ & 0.348 & 0.642 & $\downarrow$ & $<$.001 & 0.002 \\
num\_italics & $\uparrow$ & 0.003 & 0.136 & $\uparrow$ & 0.512 & 0.758 \\
bullet\_list & $\uparrow$ & 0.015 & 0.151 & $\uparrow$ & 0.239 & 0.563 \\
num\_bolds & $\uparrow$ & 0.110 & 0.399 & $\uparrow$ & 0.778 & 0.915 \\
numbered\_words & $\downarrow$ & 0.832 & 0.907 & $\uparrow$ & 0.015 & 0.088 \\
frac\_italics & $\downarrow$ & 0.253 & 0.533 & $\uparrow$ & 0.862 & 0.917 \\
reply\_frac\_in\_all & $\downarrow$ & 0.702 & 0.860 & $\downarrow$ & 0.310 & 0.589 \\
reply\_frac\_in\_content & $\downarrow$ & 0.594 & 0.792 & $\uparrow$ & 0.659 & 0.839 \\
op\_frac\_in\_stopwords & $\downarrow$ & 0.041 & 0.275 & $\downarrow$ & $<$.001 & $<$.001 \\
common\_in\_stopwords & $\uparrow$ & 0.007 & 0.146 & $\uparrow$ & 0.994 & 0.994 \\
reply\_frac\_in\_stopwords & $\downarrow$ & 0.516 & 0.737 & $\downarrow$ & 0.839 & 0.917 \\
op\_frac\_in\_all & $\uparrow$ & 0.188 & 0.501 & $\uparrow$ & $<$.001 & $<$.001 \\
jaccard\_in\_content & $\uparrow$ & 0.641 & 0.827 & $\uparrow$ & 0.725 & 0.878 \\
jaccard\_in\_stopwords & $\uparrow$ & 0.020 & 0.161 & $\uparrow$ & 0.120 & 0.361 \\
common\_in\_content & $\downarrow$ & 0.235 & 0.523 & $\uparrow$ & 0.635 & 0.839 \\
op\_frac\_in\_content & $\downarrow$ & 0.731 & 0.860 & $\downarrow$ & 0.029 & 0.134 \\
jaccard\_in\_all & $\downarrow$ & 0.068 & 0.277 & $\downarrow$ & 0.499 & 0.758 \\
\bottomrule
\end{tabular}
\end{table*}

\section{Regression of Continuous Belief Change Scores} \label{app::continuous_score}

\begin{figure}[ht]
    \centering
    \includegraphics[width=\linewidth]{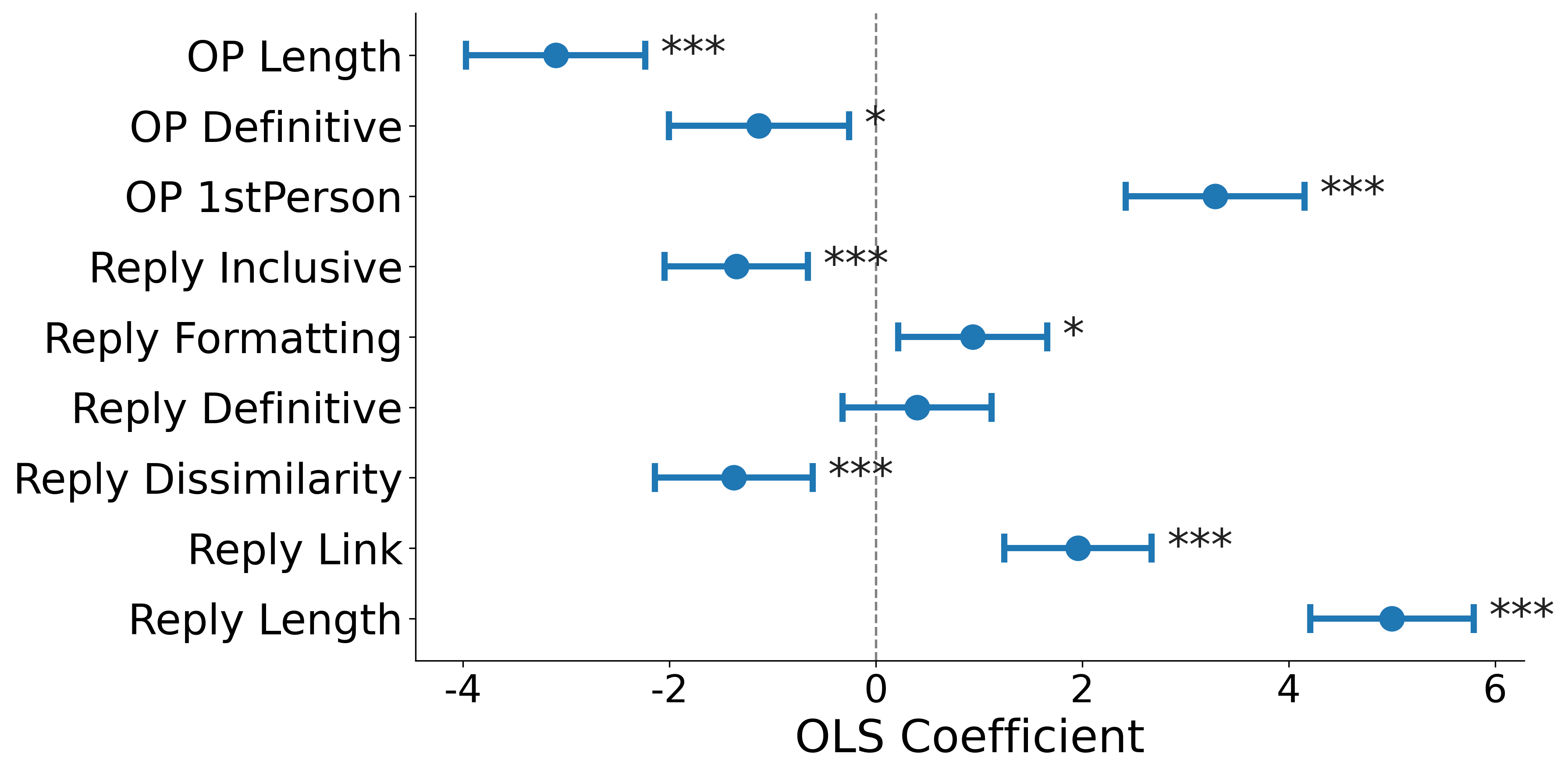}
    \caption{OLS regression coefficients of continuous belief score changes on textual features. $^{*}p<0.05$, $^{**}p<0.01$, $^{***}p<0.001$.}
    \label{fig:continuous_score}
\end{figure}

The current measure of belief change is based on the \textit{delta} signal, which is a binary measure. 
Yet in the real world, belief change can be a continuous process.
We want to test whether the continuous version of LLM belief change is also predicted by the same set of textual features.
As direct confidence measures (logprobs) are unavailable for closed-source models, we use \texttt{GPT-5.1} to score the degree of belief change reflected in each model's justification text on a 0--100 scale, yielding a continuous measure. 
To validate the continuous scores, we conducted human pairwise ranking of justification texts; the \texttt{GPT-5.1} scores achieve Kendall's $\tau = 0.889$ ($p < 0.001$) against human rankings. 
Using this score as the dependent variable, we fit an OLS regression on the same nine textual features. 
As Figure~\ref{fig:continuous_score} shows, the five largest coefficients (\textit{Reply Length}, \textit{OP 1stPerson}, \textit{Reply Link}, \textit{OP Length}, \textit{Reply Dissimilarity}) all share the same direction as in the binary logistic regression, confirming that the core feature-dependence structure holds when moving from a binary to a continuous outcome measure.

\section{Perspective Comparison} \label{app::perspective_comparison}

\begin{figure}[ht]
    \centering
    \includegraphics[width=\linewidth]{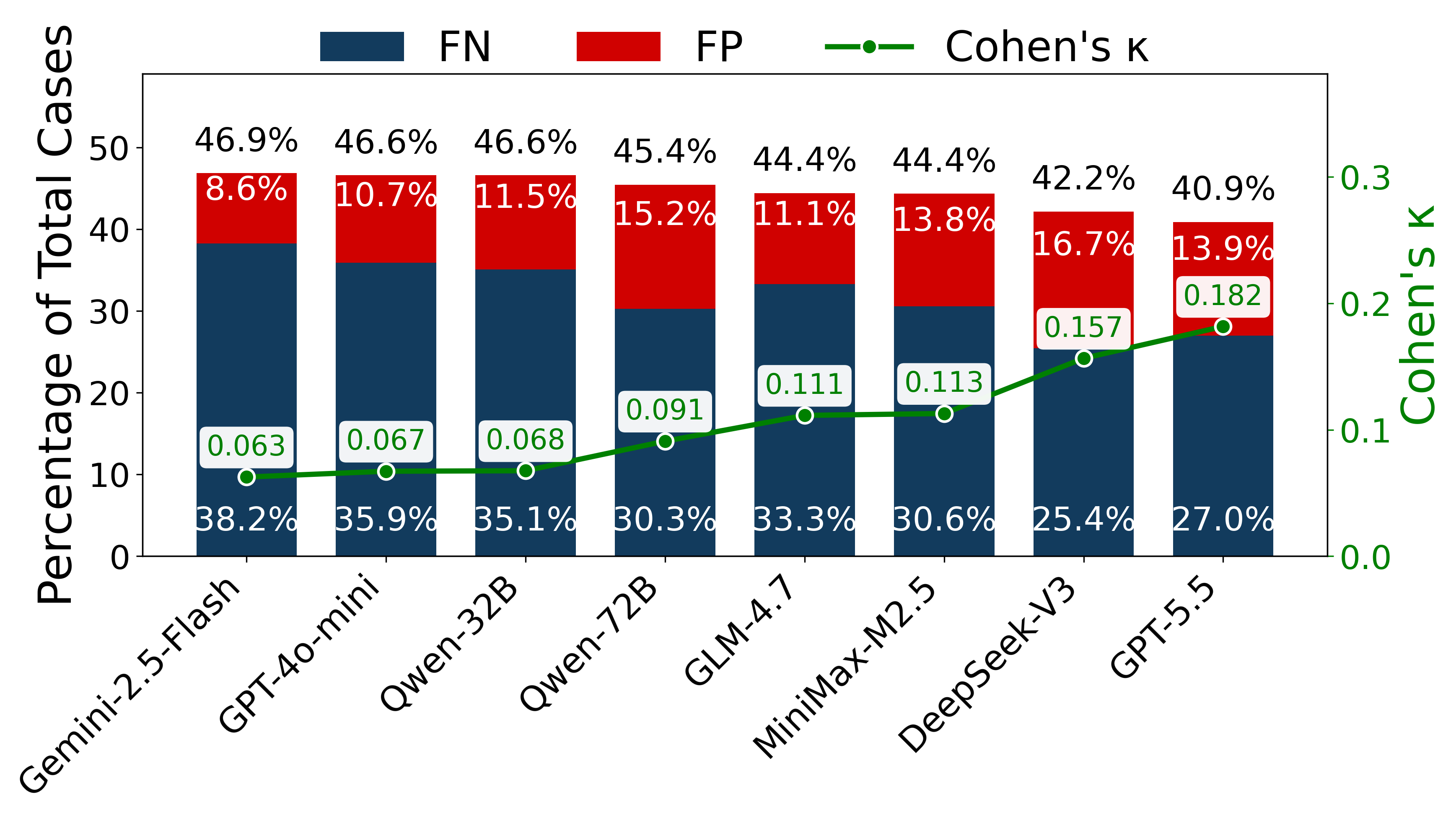}
    \caption{Discrepancy between LLMs (observer perspective) and humans in belief update judgments. LLMs are sorted by their Cohen's $\kappa$ consistency with the human counterpart.}
    \label{fig:perspective_comparison_kappa}
\end{figure}

\begin{figure}[ht]
    \centering
    \includegraphics[width=\linewidth]{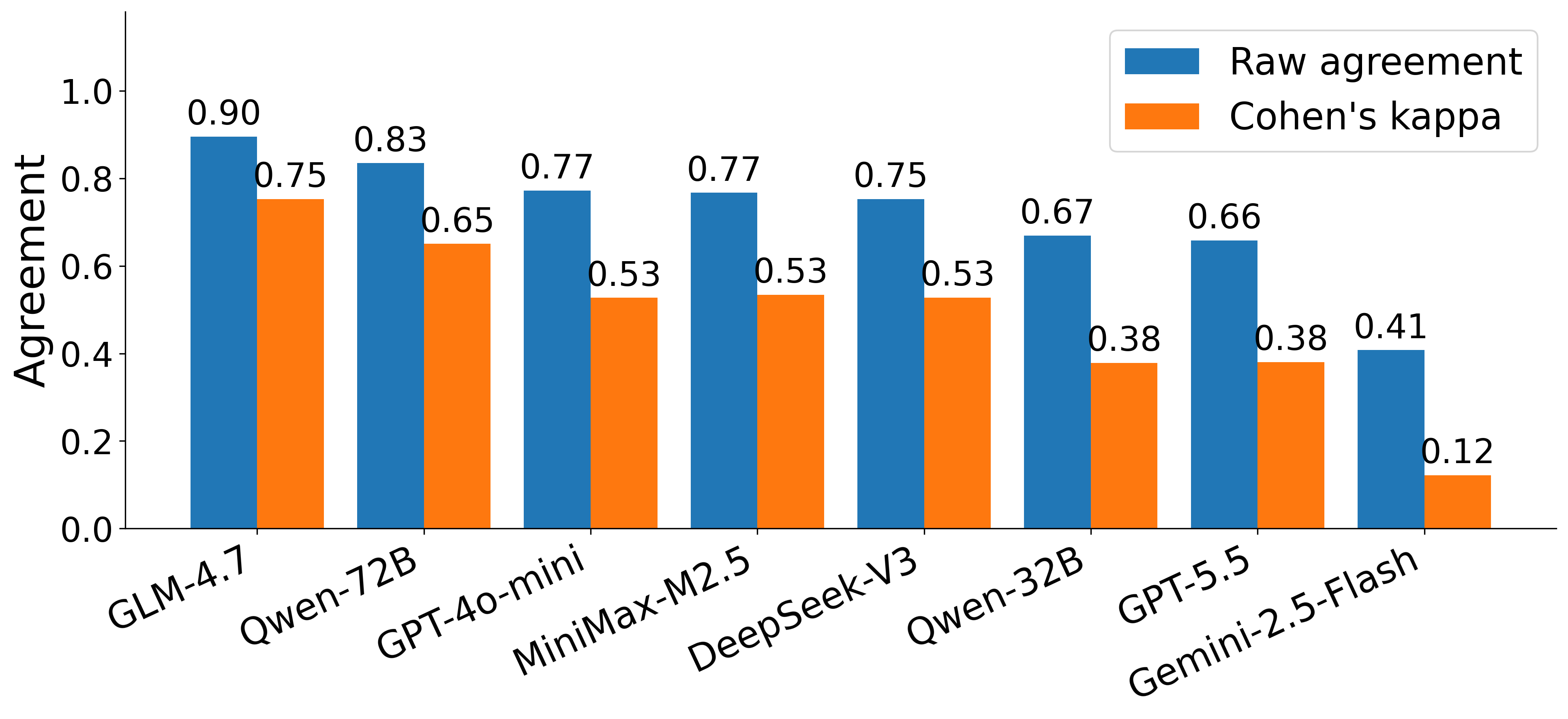}
    \caption{Judgment consistency between first-person and observer perspectives. LLMs are sorted by their raw agreement between two perspectives.}
    \label{fig:perspective_consistency}
\end{figure}

Figure~\ref{fig:perspective_comparison_kappa} presents the error composition and Cohen's $\kappa$ for all eight models under the observer condition, serving as the observer-side counterpart to Figure~\ref{fig:anomalies_and_kappa}.
Figure~\ref{fig:perspective_consistency} reports the judgment consistency between each model's first-person and observer outputs, measured by both raw agreement and Cohen's $\kappa$.
\texttt{GLM-4.7} exhibits the highest cross-perspective consistency (raw agreement = 0.90, $\kappa$ = 0.75), while \texttt{Gemini-2.5-Flash} exhibits the lowest (raw agreement = 0.41, $\kappa$ = 0.12). 
This ordering aligns with the magnitude of error-profile reversal observed in Section~\ref{sec::RQ3}: \texttt{Gemini-2.5-Flash} undergoes the most dramatic shift from a receptive profile under the first-person condition to a resistant profile under the observer condition, whereas \texttt{GLM-4.7}, already resistant under the first-person condition, changes comparatively little.

\section{AI Assistant Usage Disclosure}

We used AI assistants (Claude, ChatGPT, Gemini) to assist with coding as well as drafting, editing, and proofreading portions of the manuscript. 
All research design, experimental decisions, data analysis, interpretation, and argumentation are the authors' own.

\end{document}